\documentclass[twocolumn,trackchanges,openany]{aastex631}

\usepackage{natbib}
\usepackage{graphicx}
\usepackage{epstopdf}
\usepackage{amsmath}
\usepackage{multirow}   
\usepackage{booktabs}
\usepackage{float}
\usepackage{placeins}

\let\oldAA\AA
\renewcommand{\AA}{\text{\normalfont\oldAA}}

\newcommand{\sersic}{S\'{e}rsic}
\newcommand{\OIII}{[O\,\textsc{iii}]}
\newcommand{\NII}{[N\,\textsc{ii}]}
\newcommand{\CII}{[C\,\textsc{ii}]}
\newcommand{\Ha}{H$\alpha$}
\newcommand{\Hb}{H$\beta$}
\newcommand{\sfrha}{$\rm SFR_{\rm H\alpha}$}
\newcommand{\sfruv}{$\rm SFR_{\rm NUV}$}
\newcommand{\starmass}{$M_{\rm *}$}

\begin{document}

\title{JWST Insights into Narrow-line Little Red Dots}

\author[0000-0002-2420-5022]{Zijian Zhang}
\affiliation{Kavli Institute for Astronomy and Astrophysics, Peking University, Beijing 100871, China}
\affiliation{Department of Astronomy, School of Physics, Peking University, Beijing 100871, China}

\author[0000-0003-4176-6486]{Linhua Jiang}
\affiliation{Kavli Institute for Astronomy and Astrophysics, Peking University, Beijing 100871, China}
\affiliation{Department of Astronomy, School of Physics, Peking University, Beijing 100871, China}

\author[0000-0002-4385-0270]{Weiyang Liu}
\affiliation{Kavli Institute for Astronomy and Astrophysics, Peking University, Beijing 100871, China}
\affiliation{Department of Astronomy, School of Physics, Peking University, Beijing 100871, China}

\author[0000-0001-6947-5846]{Luis C. Ho}
\affiliation{Kavli Institute for Astronomy and Astrophysics, Peking University, Beijing 100871, China}
\affiliation{Department of Astronomy, School of Physics, Peking University, Beijing 100871, China}

\author[0000-0001-9840-4959]{Kohei Inayoshi}
\affiliation{Kavli Institute for Astronomy and Astrophysics, Peking University, Beijing 100871, China}

\begin{abstract}
JWST has revealed a population of red and compact objects with unique V-shape SEDs known as ``Little Red Dots'' (LRDs). Many LRDs exhibit broad Balmer lines and thus likely host AGNs. Here we present a study of five LRDs without broad \Ha\ lines at $z\ge 5$. They are selected from 32 LRDs that have NIRSpec high- or medium-resolution grating spectra covering \Ha, and their \Ha\ line widths are around 250 $\rm km~s^{-1}$. Compared to normal star-forming galaxies, narrow-line LRDs tend to have relatively higher \Ha\ line widths and luminosities. If they are dominated by galaxies, our SED modeling suggests that they are dusty, compact star-forming galaxies with high stellar masses and star formation rates. Alternatively, if their SEDs are produced by AGNs, the inferred black hole masses ($10^5$–$10^6~\rm M_\odot$) place them at the low-mass end of the AGN population. They may represent an early stage of super-Eddington growth, where the black holes have yet to accumulate significant masses. With large uncertainties, these black holes appear slightly over-massive relative to the local $M_{\rm BH}$–$M_{\rm *}$ relation, but consistent or under-massive with respect to the $M_{\rm BH}$–$\sigma_{\rm stellar}$ and $M_{\rm BH}$–$M_{\rm dyn}$ relations. These narrow-line LRDs offer new insights into the diversity of LRDs. Finally, we find that $\sim44\%$ of high-redshift broad-line AGNs exhibit V-shape SEDs. In addition, $\sim20\%$ of the previous LRD candidates are not real LRDs, and their V-shape SEDs are caused by strong line emission.

\end{abstract}

\section{Introduction}
\label{sec:intro}

One of the most intriguing results from early observations with the James Webb Space Telescope (JWST) is the identification of an abundance of red and compact objects \citep[e.g.,][]{2023Natur.616..266L,2024ApJ...963..128B}. These objects exhibit a unique ``V-shape'' spectral energy distribution (SED) that is red in the rest-frame optical and blue in the rest-frame ultraviolet (UV). They are called ``Little Red Dots'' (LRDs) for their compact morphology \citep[typical $R_{\rm eff} \sim 100$ pc; e.g.,][]{2023ApJ...955L..12B,2023ApJ...952..142F,2024RNAAS...8..207G} and red colors in the observed-frame $\sim 2$--$5 ~\mu m$ range \citep[e.g.,][]{2024ApJ...963..129M}. LRDs are ubiquitous at redshifts from $z \sim 4$ up to $z \sim 9$ \citep[e.g.,][]{2024ApJ...963..128B,2023ApJ...954L..46L,2024arXiv240403576K,2024ApJ...968...38K}, while at lower redshift their number density decline rapidly \citep[e.g.,][]{2024arXiv240403576K,2025arXiv250408032M}. Spectroscopy of more than one hundred photometrically identified LRDs reveals that $\gtrsim 70\%$ of them exhibit broad Balmer lines \citep[e.g.,][]{2023ApJ...959...39H,2024ApJ...964...39G,2024A&A...691A..52K,2024arXiv240403576K,2024ApJ...963..129M,2024arXiv240302304W}. 

The physical origins of the V-shape SED and the nature of LRDs remain unclear. A natural explanation of the broad lines and compact sizes is that LRDs are active galactic nuclei (AGNs). However, the presence of prominent Balmer breaks observed in several LRDs \citep[e.g.,][]{2024arXiv240807745B,2024arXiv240720320K,2024arXiv240302304W}, along with the general lack of X-ray \citep[e.g.,][]{2024ApJ...969L..18A,2024ApJ...974L..26Y,2025MNRAS.538.1921M}, mid-IR emission \citep[e.g.,][]{2024ApJ...968....4P,2024ApJ...968...34W}, and variability signatures \citep[e.g.,][]{2024arXiv240704777K,2025ApJ...983L..26T,2025ApJ...985..119Z}, pose challenges to the AGN interpretation. These issues, however, may be mitigated under a scenario involving super-Eddington accreting black holes enshrouded by extremely dense gas \citep[e.g.,][]{2024arXiv241203653I,2025ApJ...980L..27I,2025arXiv250113082J,2025arXiv250316596N}, supported by the high fraction ($\sim 20\%$) of Balmer absorption in high- and medium-resolution spectra \citep[e.g.,][]{2024arXiv240717570L,2024arXiv240906772T,2025arXiv250403551J}. In this context, LRDs may represent the very first activity of supermassive black hole (SMBH) growth \citep{2025arXiv250305537I}, potentially linked to the formation of seed black holes via non-standard seeding channels, such as the core-collapse of self-interacting dark matter halos \citep{2025arXiv250323710J}. 

While the majority of spectroscopically confirmed LRDs exhibit broad emission lines, it remains unclear whether the characteristic V-shape continuum is necessarily associated with broad-line emission. Some broad emission lines of LRDs were identified using low-resolution ($R \sim 100$) NIRSpec/PRISM spectra \citep[e.g.,][]{2024ApJ...964...39G}, and their intrinsic line widths can be much smaller. For instance, one LRD with a broad H$\alpha$ FWHM $ = 1900 \pm 210~ \rm km~s^{-1}$ reported by \citet{2024ApJ...964...39G} was later found to be well fitted by a single Gaussian with FWHM $ =251 \pm 39 \rm ~km~s^{-1}$ in higher-resolution ($R \sim 1600$) NIRCam/WFSS slitless spectra \citep{2025arXiv250303829F}. More recently, \citet{2025arXiv250502895Z} report that 68\% of color-selected LRDs with \Ha\ detections in the NIRCam/Grism spectra in the GOODS-N field do not exhibit broad-line features. Since the detection of broad components in these slitless spectra is limited by sensitivity, deeper and higher-resolution NIRSpec data are needed to confirm their absence. Nevertheless, these results suggest that a subset of LRDs may exhibit only narrow lines while sharing key photometric properties with their broad-line counterparts, such as compact sizes and V-shape SEDs. Understanding the physical origins of narrow-line LRDs may provide crucial insights into the formation pathways and duty cycles of LRDs. 

The situation is further complicated by the lack of a clear definition and robust selection methodology for LRDs in the literature. A commonly adopted definition identifies LRDs as sources with compact morphologies at long wavelengths and V-shape SEDs. Some studies further require the presence of broad emission lines as part of the definition. In practice, many sources are classified as LRDs based on simple color-cut or slope-cut criteria \citep[e.g.,][]{2024arXiv240610341A,2024arXiv240403576K,2024ApJ...968...38K,2025ApJ...978...92L}. These photometric colors and fitted slopes may be contaminated by strong emission lines, such as \Ha, \Hb, and \OIII\ line \citep[e.g.,][]{2025ApJ...979..138H}. This leads to misclassifications or the inclusion of unrelated sources in LRD samples. Therefore, it is crucial to remove the contamination of the line emission when constructing a LRD sample. 

In this work, we present a systematic investigation of narrow-line LRDs based on a clean sample selected using emission-line-free photometry. The contamination from strong emission lines is removed using spectroscopic information from NIRSpec/PRISM observations. The narrow \Ha\ line is identified with high- or medium-resolution NIRSpec grating spectra. We compare the properties of narrow-line LRDs with those of broad-line LRDs and normal star-forming galaxies, aiming to understand the diversity within the LRD population and the possible links between them. The paper is organized as follows. In Section \ref{sec:data_reduction}, we describe our data and data reduction. In Section \ref{sec:LRD_narrowline_selection}, we introduce our selection and characterization of the narrow-line LRD sample. Two possible interpretations of the narrow-line LRDs are proposed in Section \ref{sec:two_scenario}. In Section \ref{sec:discussion}, we discuss the comparison of the sample selection methods and the implication of narrow-line LRDs. Conclusions are given in Section \ref{sec:conclusion}. Throughout this paper, we use a cosmology with $H_{0}=67.4~ \rm km~s^{-1}~Mpc^{-1}$, $\Omega_{\rm M} = 0.315$, and $\Omega_{\rm \Lambda} = 0.686$ \citep{2020A&A...641A...6P}. By default, measurement uncertainties are quoted at a 1$\sigma$ confidence level, and upper limits are quoted at a 90\% confidence level.

\section{Data Description and Reduction}
\label{sec:data_reduction}

\subsection{Imaging Data and Catalog}
\label{subsec:image_data}

We collect publicly available JWST/NIRCam data in A2744, CEERS, GOODS-S, GOODS-N, COSMOS, and UDS fields. The detailed reduction steps can be found in \citet{2025ApJ...985..119Z}. Briefly, we reduce the raw data using the combination of the \textit{JWST Calibration Pipeline} \citep[v.1.12.5;][]{2023zndo..10022973B}, the CEERS NIRCam imaging reduction \citep{2023ApJ...946L..12B},\footnote{\url{https://github.com/ceers/ceers-nircam.}} and our own custom codes. All NIRCam images are aligned to the HST images in each field. We use the public HST image from the CANDELS and 3D-HST surveys \citep{2011ApJS..197...35G,2011ApJS..197...36K,2012ApJS..200...13B,2016ApJS..225...27M}. The final mosaics in all fields have pixel scales of $0\farcs03$.

The empirical PSF models of each field are constructed from the mosaic images with \texttt{PSFEx} \citep{2011ASPC..442..435B}, following the method in \citet{2024ApJ...962..139Z}. We then use these PSFs to derive kernels to match all PSFs to the F444W PSF using \texttt{PyPHER} \citep{2016ascl.soft09022B}. These kernels are used to convolve all images to match the F444W image. Then we perform photometry using \texttt{SExtractor} \citep{1996A&AS..117..393B} on the PSF-matched images to construct source catalog. We use the dual image mode and the detection image is an $\chi^2$ image combining the PSF-matched F277W, F356W and F444W images. We adopt a two-step `cold + hot' mode to achieve an optimal source detection \citep{2012MNRAS.422..449B}. Photometry is measured in a small Kron aperture ($k$=1.1, $R_{\min}$=1.6), and an aperture correction is performed to match the default Kron aperture ($k$=2.5, $R_{\min}$=3.5) in each band. Following the method in \citet{2023ApJS..269...16R}, we measure the photometric errors by placing 100,000 random apertures in the source-free region of the images. These apertures are divided into 100 groups based on their weight values. For each group, a power-law relation between aperture size and flux scatter is fitted and used to derive the sky background noise for each source according to its weight value and aperture size. 

For the CEERS field, we adopt the ASTRODEEP catalog \citep{2024A&A...691A.240M} that is based on the images from the CEERS data release \citep{2023ApJ...946L..12B}, since this catalog includes more HST bands. Their basic procedure for photometry is similar to ours, with the major differences outlined below. Source detection was carried out using SExtractor on a weighted combination of the F356W and F444W images. The total fluxes were inferred within a preferred aperture, determined via the object’s segmentation map. The aperture correction in the detection band was applied to other bands under the assumption of no color gradient.

\subsection{NIRSpec Spectroscopic Data}

The NIRSpec spectroscopic data used in this work are from the DAWN JWST Archive (DJA\footnote{\url{https://dawn-cph.github.io/dja/}.}), which includes fully reduced 1D and 2D spectra of all public JWST/NIRSpec data. We use the version 3 publicly released NIRSpec datasets, including 26,077 spectra for 14,909 sources from 43 programs. The source IDs used in this paper are the UIDs of the DJA dataset. These data are reduced using the latest version of \texttt{msaexp} and the \textit{JWST Calibration Pipeline} (v.1.14.0) with calibration files \texttt{jwst\_1225.pmap} from the Calibration Reference Data System. The detailed MSA data reduction procedures using \texttt{msaexp} can be found in \citet{deGraaff_2025_RUBIES} and \citet{2024Sci...384..890H}. The absolute flux calibration accuracy for NIRSpec MSA spectroscopy is generally at a level of 10\% to 20\% \citep{deGraaff_2025_RUBIES}. DJA also offers spectroscopic redshifts obtained through the template fitting algorithms employed in the \texttt{msaexp} pipeline. These redshifts are visually inspected and graded for reliability, with grade 3 indicating a secure measurement. In our analysis, we only include sources with grade 3 redshift, and adopt the grade 3 redshift as their redshift value.

We also notice the availability of two large NIRSpec datasets from the data releases of two JWST Cycle 1 programs: JWST Advanced Deep Extragalactic Survey \citep[JADES;][]{2023arXiv230602465E,d2024jades} and the CEERS \citep{2023ApJ...946L..12B}. These datasets are carefully reduced and generally offer higher data quality compared to the DJA data \citep[e.g.,][]{2025ApJ...980...93J}. For uniformity, we adopt only the DJA data throughout this work. We have verified that this choice has a negligible effect on our results. Most of the CEERS spectra are also included in DJA. In the JADES release, 1092 out of 2525 targets are not available in DJA; however, we identified only 4 LRDs (and no narrow-line LRDs) among them, which may be attributed to the target selection strategy of the JADES survey. Overall, the results obtained from the JADES/CEERS and DJA datasets are consistent for common targets.

\section{Narrow-line LRDs}
\label{sec:LRD_narrowline_selection}

\subsection{Selection of LRDs}
\label{subsec:LRD_selection}

Photometric selection of LRDs in the literature primarily relies on their observed colors to identify the characteristic V-shape SEDs \citep[e.g.,][]{2024arXiv240610341A,2024ApJ...968...38K,2025ApJ...978...92L}. \citet{2024arXiv240403576K} further refined this approach by applying cuts to the rest-frame optical and UV slopes. This photometry-only selection may suffer from inaccurate photo-$z$ or boosted red colors by strong emission lines \citep[e.g.,][]{2025ApJ...979..138H}.

\begin{figure}
\centering
\begin{minipage}{0.96\linewidth}
  \includegraphics[width=\textwidth]{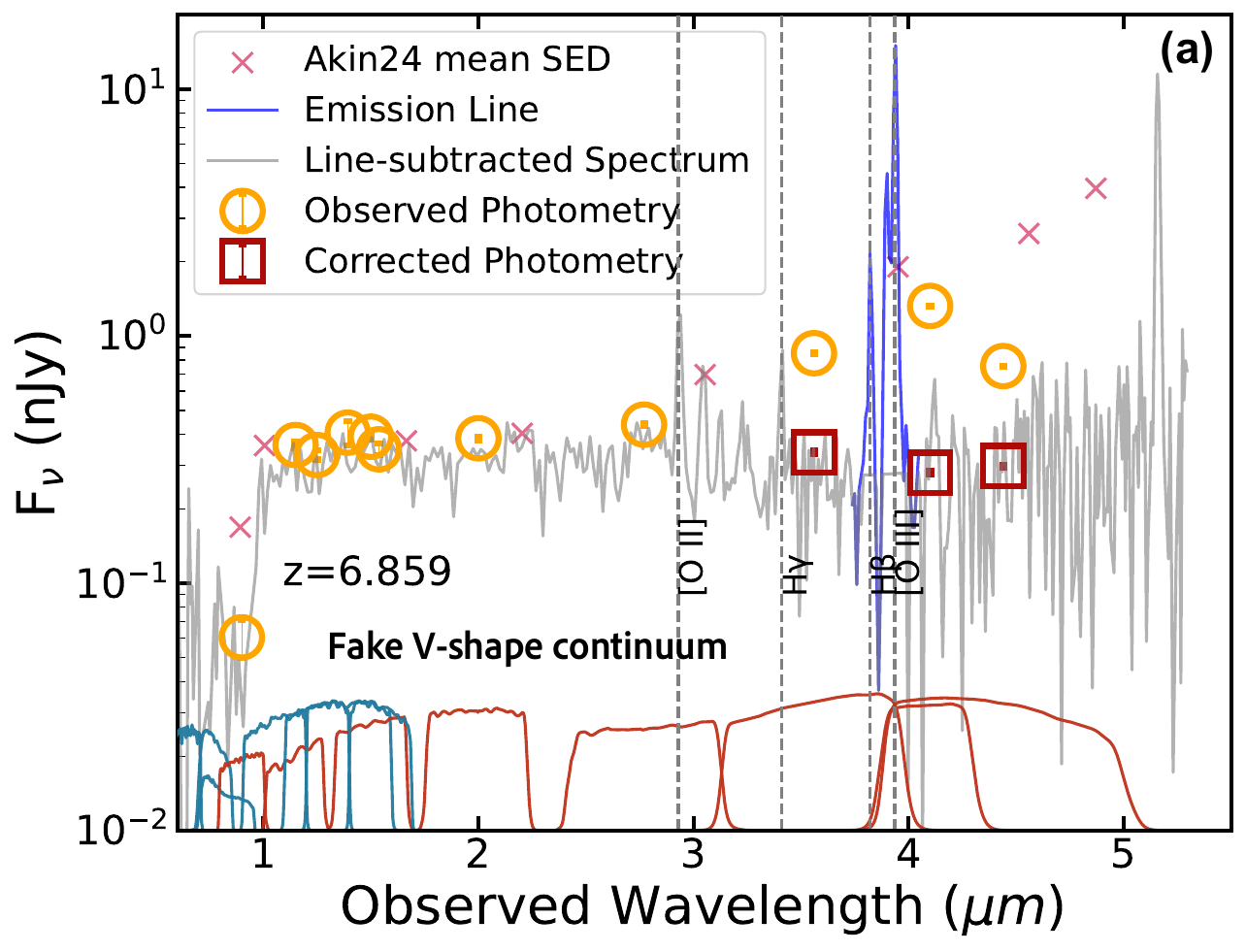}
\end{minipage}
\vspace{0.5em} 
\begin{minipage}{0.96\linewidth}
  \includegraphics[width=\textwidth]{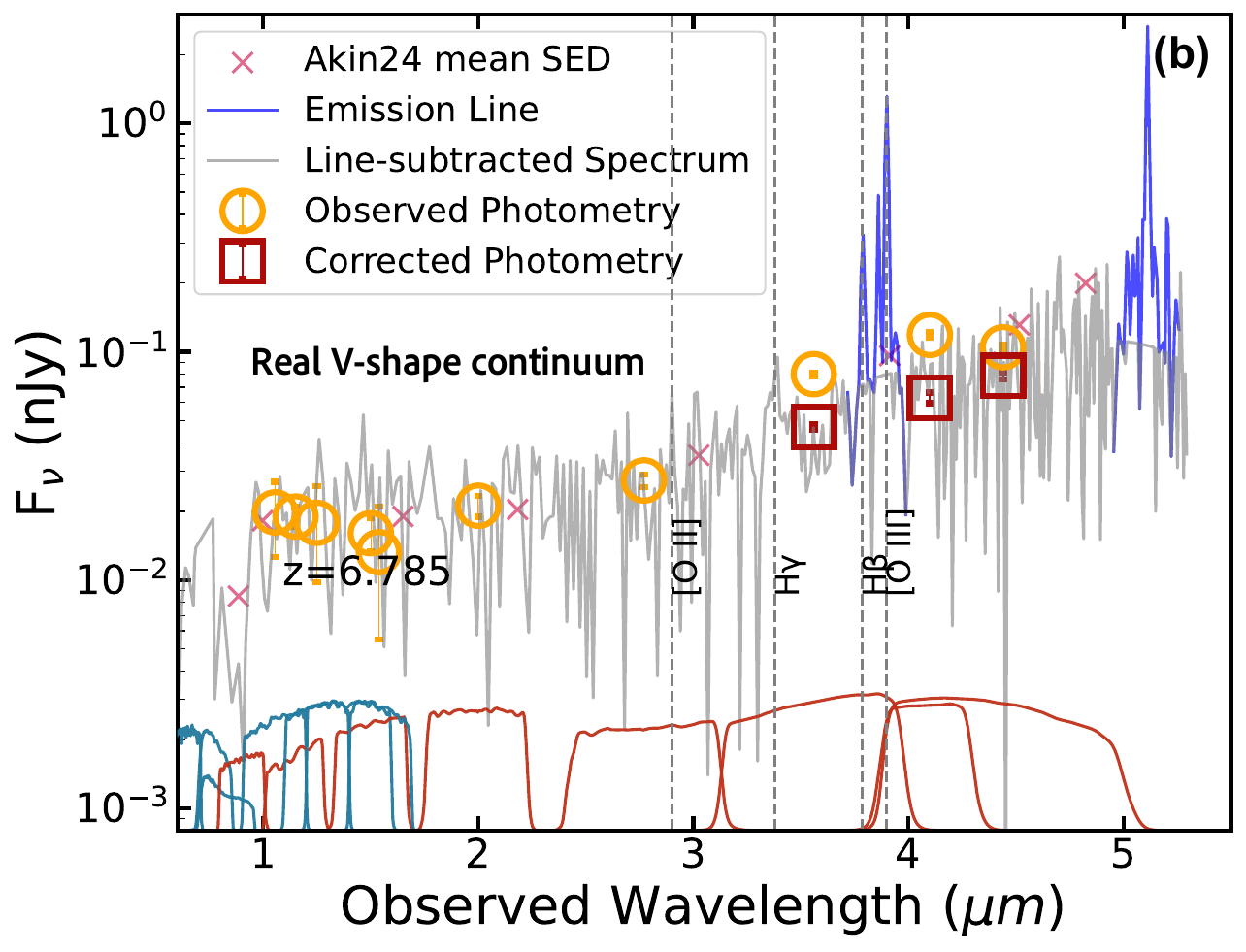}
\end{minipage}
\caption{Example of sources with a fake V-shape continuum (a) and a real V-shape continuum (b). In each panel, we show the line-subtracted spectrum in gray and the emission lines (\Hb, \OIII, and \Ha\ lines) in blue. The observed photometry and line-free photometry are shown as the orange circles and red squares, respectively. The average SED of LRDs from \citet{2024arXiv240610341A} are shown as the pink cross for comparison. The redshift and main emission lines are labeled. The JWST and HST transmission curves are in red and teal blue, respectively.}
\label{fig:slope_fitting_demo}
\end{figure}

In order to select reliable LRDs, we begin with a parent sample of 5444 sources that have NIRSpec/PRISM spectra with $z>2$. These spectra provide accurate redshifts and help us eliminate the emission line contribution. We first calculate the flux contribution of the \Hb, \OIII, and \Ha\ emission lines in the relevant photometric bands using the spectra as follows. Other lines are much weaker and thus not considered. We replace the strong line in the original spectrum with a continuum extrapolated from the nearby continuum to construct a line-free spectrum. The original and line-free spectra are convolved with the filter transmission curve of a broad (or medium) band separately, and the ratio of the resultant values indicates the pure continuum fraction in this band. Then this ratio is applied to the broad (or medium) band photometry to obtain a pure continuum photometry. We use the pure continuum photometry for each source to calculate continuum slopes. 
The rest-frame UV slope is fitted using bands with rest-frame pivot wavelengths $\lambda_{\rm pivot,rest} \in [1350, 3645]\AA$, while the rest-frame optical slope is fitted with bands with $\lambda_{\rm pivot,rest} \in [3300, 8000]\AA$. The lower bound of the optical range is set slightly below the Balmer break to accommodate high-redshift sources, which often have only two or three bands available at the red end. Including a bluer band in such cases improves the robustness of the slope fitting and has negligible impact on low-redshift sources. 

\begin{figure}
 \includegraphics[width=0.44\textwidth]{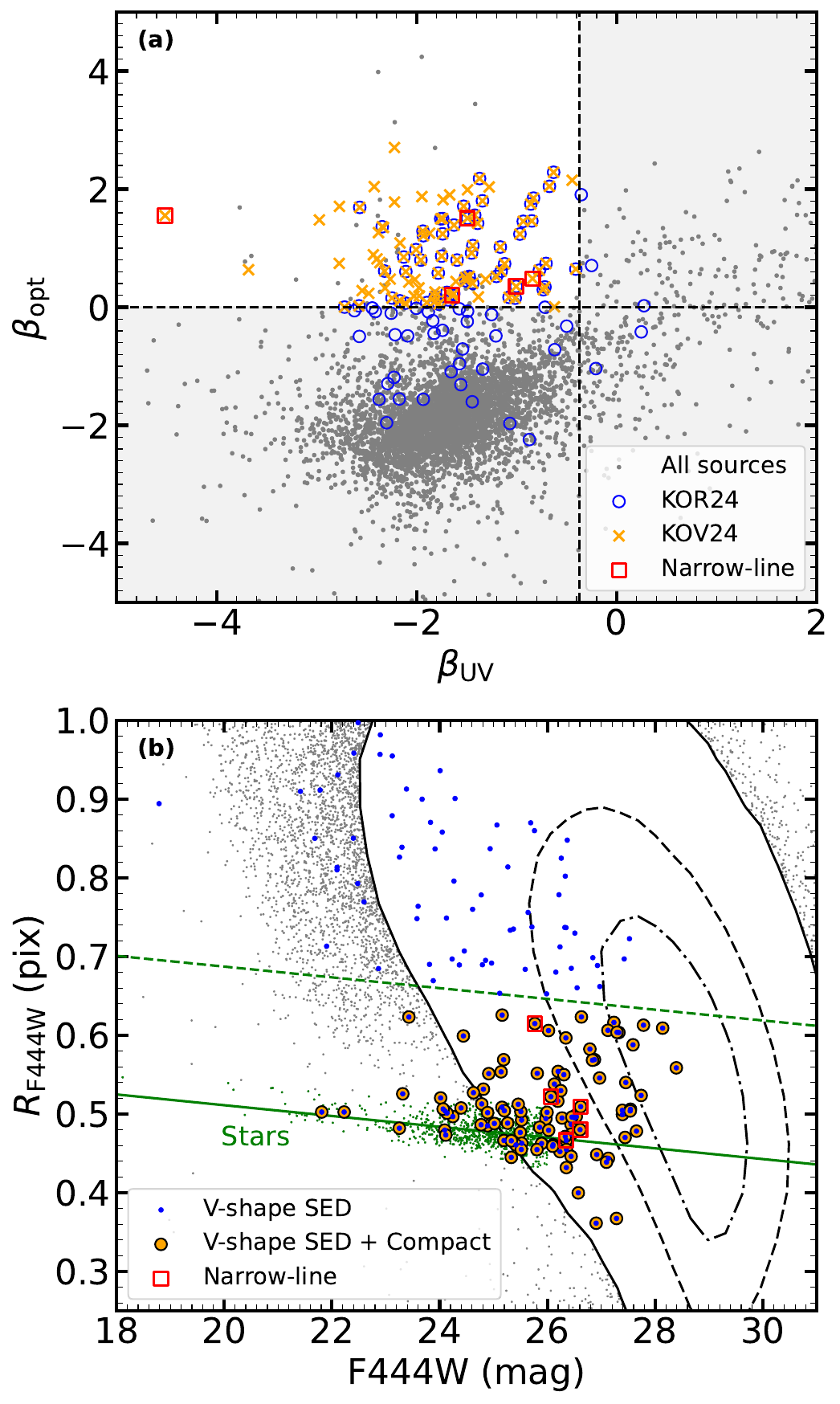}
 \centering
 \caption{(a) Distribution of the best-fit UV and optical slopes using the line-free photometry. These sources represent all $z>2$ sources with NIRSpec/PRISM spectra and $\rm S/N>10$ in F444W. The blue circles and yellow crosses are sources that satisfied the LRD selection criteria of \citet{2024ApJ...968...38K} and \citet{2024arXiv240403576K}, respectively. The red squares represent the narrow-line LRDs. (b) F444W magnitude versus half-light radius for sources that satisfy the LRD selection criteria of \citet{2024arXiv240403576K}. Green circles are stars and the solid line denotes the best-fit to the stellar locus. The magnitude-dependent size cut is shown as dashed line.}
 \label{fig:selection}
\end{figure}

LRDs are then selected using the criteria of $\beta_{\rm opt} > 0$ and $\beta_{\rm UV} < -0.37$, as done by \citet{2024arXiv240403576K}. We present two examples in Figure \ref{fig:slope_fitting_demo} illustrating sources with a fake and a genuine V-shape continuum, respectively. For comparison, we also show the LRD template constructed from the average SED of LRDs in \citet{2024arXiv240610341A}. In Figure \ref{fig:slope_fitting_demo}(a), the red color of the source is primarily due to its strong \Hb\ and \OIII\ emission lines. After subtracting their contributions, the optical slope becomes flat. In contrast, the source shown in Figure \ref{fig:slope_fitting_demo}(b) remains its V-shape after the emission line contribute is subtracted. The distribution of the best-fit UV and optical slopes for our sample is shown in Figure \ref{fig:selection}(a). In total, there are 179 sources with $\beta_{\rm opt} > 0$ and $\beta_{\rm UV} < -0.37$.

\begin{figure}
 \includegraphics[width=0.44\textwidth]{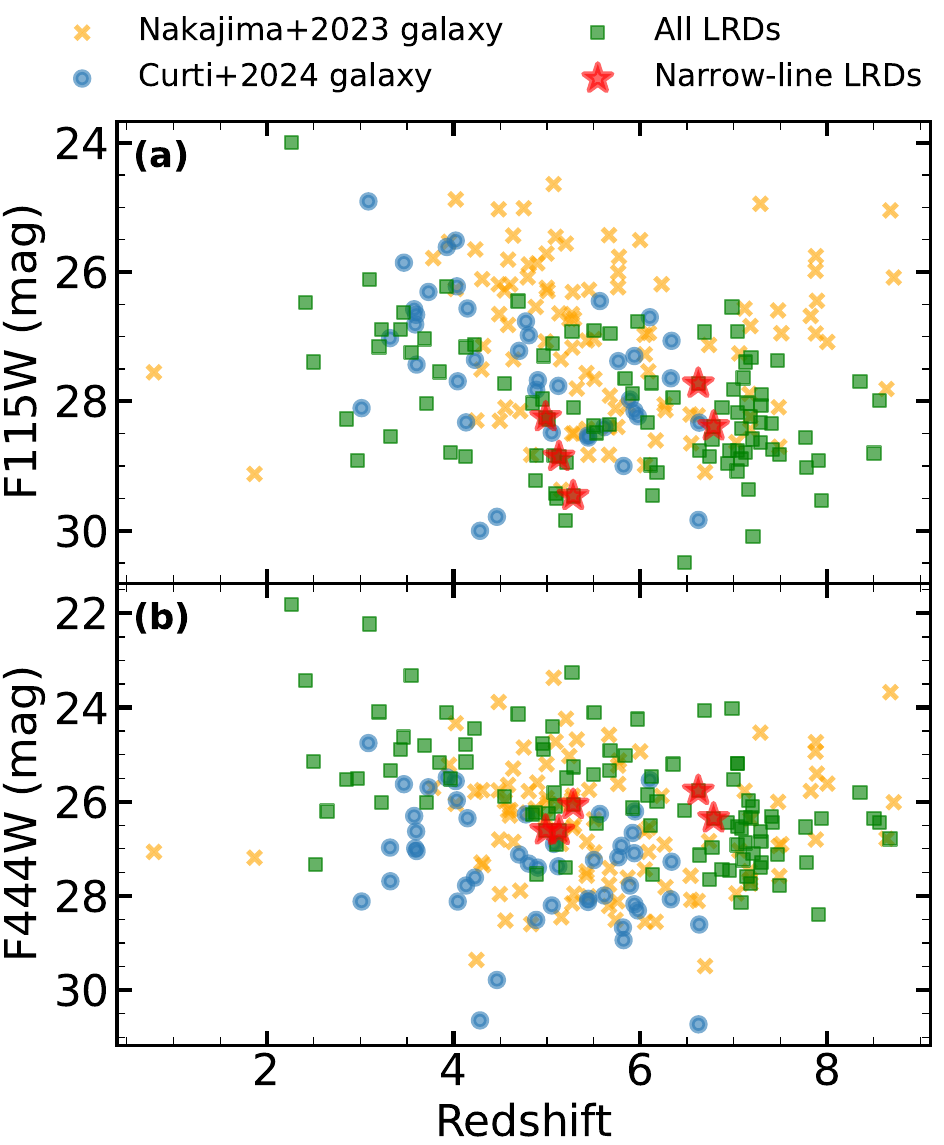}
 \centering
 \caption{(a) Source distribution in the redshift-magnitude diagram. The five narrow-line LRDs are enclosed in the red stars. All LRDs identified in this work are shown as the green squares. The star-forming galaxies from \citet[yellow crosses]{2023ApJS..269...33N} and \citet[blue dots]{2024A&A...684A..75C} are shown for comparison.}
 \label{fig:z_mag}
\end{figure}

To select sources with compact morphology, we apply an additional cut on the half-light radius $R_{\rm F444W}$ measured by SExtractor in the F444W-band images. We derive a relation between $m_{\rm F444W}$ and $R_{\rm F444W}$ for point sources, as shown in Figure \ref{fig:selection}(b). Sources with $R_{\rm F444W}$ below 1.5 times the relation (green dashed line in Figure \ref{fig:selection}b) are selected. A total of 98 LRDs meet this criterion. The redshift-magnitude diagrams (F115W and F444W) of the selected LRDs are shown in Figure \ref{fig:z_mag}. These LRDs span a large redshift range of $z \sim 2$--9 and a broad magnitude range of $m_{\rm F444W} \sim 22$--28 mag. In the rest-frame UV, their magnitudes are comparable to those of typical star-forming galaxies reported by \citet{2024A&A...684A..75C}, whereas in the rest-frame optical, the LRDs appear brighter.

\begin{table*}
\setlength{\tabcolsep}{2pt}
\caption{Basic Properties for the Five Narrow-line LRDs}
\label{tab:base_info}
\hspace*{-2.5cm}
\begin{tabular}{ccccccccccccc} 
\hline
\hline
UID & RA & DEC & $z$ & $\beta_{\rm UV}$ & $\beta_{\rm opt}$ & $\rm BIC_{H\alpha2}$ & $\rm BIC_{H\alpha1}$ & $\rm FWHM_{H\alpha}$ & $L_{\rm H\alpha,obs}$    & $\rm EW_{H\alpha}$      & $L_{\rm H\beta,obs}$      & $\rm FWHM_{[O\,\textsc{iii}]}$  \\ 
& & & & & & & & $\rm km~s^{-1}$ & $10^{41} \rm~erg~s^{-1}$ & $\textbf{\rm \AA}$ & $10^{41} \rm~erg~s^{-1}$ & $\rm km~s^{-1}$  \\
(1)&(2)&(3)&(4)&(5)&(6)&(7)&(8)&(9)&(10)&(11)&(12)&(13)\\
\hline
8219       & 215.078259 & 52.948497 & 6.785 & $-1.64$    & 0.20      & 72.6   & 42.9   & $136\pm36$       & $16.4\pm2.9$ & $109.3\pm19.1$ & $3.2\pm0.8$ & $196\pm22$          \\ 
\hline
12329      & 214.797537 & 52.818746 & 6.620 & $-1.00$    & 0.35      & 41.9   & 38.5   & $292\pm 35$       & $20.5\pm3.0$ & $179.3\pm26.3$ & $3.4\pm1.0$  & $117\pm90$           \\ 
\hline
16321      & 34.324122  & $-5.252789$ & 4.985 & $-1.49$    & 1.51      & 71.8   & 68.0   & $475_{-74}^{+42}$       & $4.3\pm1.0$ & $81.0\pm 18.9$  & $<3.9$ & --                       \\ 
\hline
20547      & 34.455376  & $-5.231814$ & 5.278 & $-4.51$    & 1.55      & 66.2   & 56.9    & $272\pm64$        & $3.5\pm0.9$ & $72.3\pm18.6$  & $1.5\pm1.2$ & $124\pm111$          \\ 
\hline
22015      & 34.374854  & $-5.275528$ & 5.127  & $-0.84$    & 0.48      &  49.8  & 48.8   & $229\pm36$        & $5.8\pm0.9$ & $179.9\pm 28.0$ & $2.1\pm1.1$ & $94\pm87$            \\
\hline
\hline
\end{tabular}
\tablecomments{
Col. (1): UID in the DJA catalog.
Cols. (2) and (3): Right Ascension and Declination.
Col. (4): Redshift.
Cols. (5) and (6): Rest-frame UV and optical continuum slopes.
Cols. (7) and (8): BIC values of the fitting using broad+narrow Gaussian components and only a narrow Gaussian component. 
Col. (9): Intrinsic \Ha\ FWHM.
Col. (10): Observed \Ha\ luminosity.
Col. (11): \Ha\ equivalent width.
Col. (12): Observed \Hb\ luminosity.
Col. (13): Intrinsic \OIII\ FWHM.
}
\end{table*}

\subsection{Selection of Narrow-line LRDs}
\label{subsec:emission_line_fitting}

\begin{figure*}
 \includegraphics[width=0.99\textwidth, trim=0 0 10 0]
 {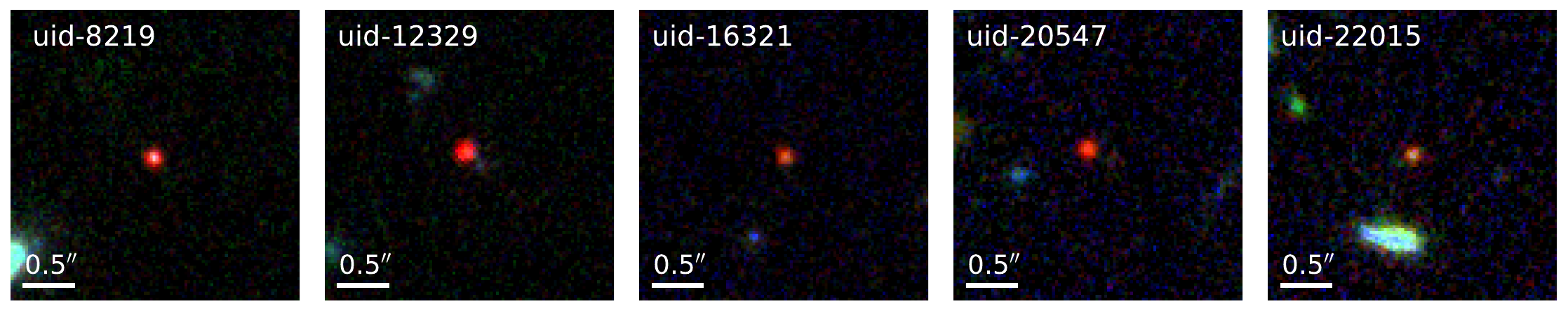}
 \includegraphics[width=1.0\textwidth, trim=35 0 0 0]{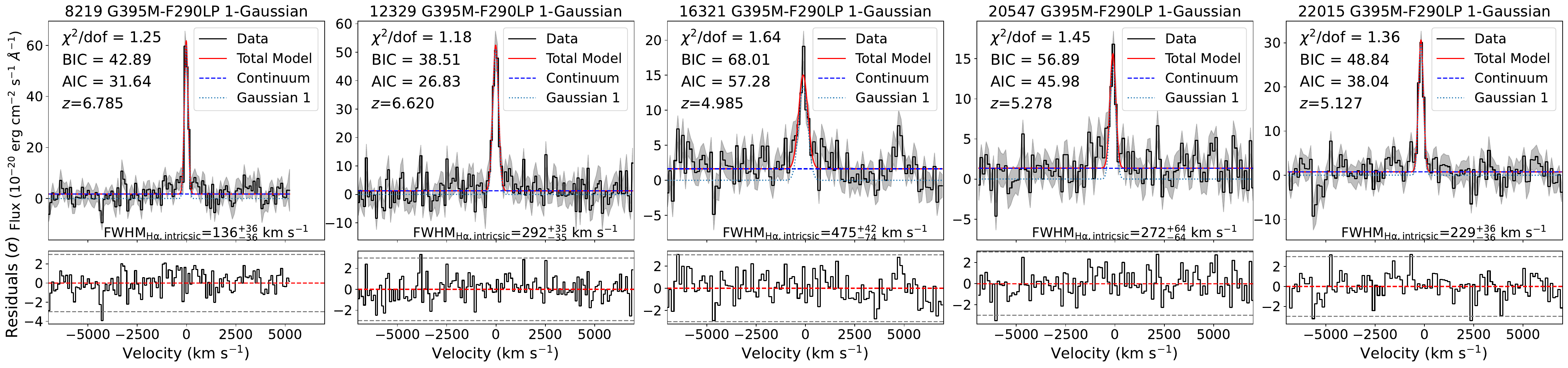}
 \centering
 \caption{(a) NIRCam images of the five narrow-line LRDs. The following false-color coding was adopted: blue - F115W; green - F200W; red - F444W. For uid-16321, uid-20547, and uid-22015 in the UDS field, the F200W band is not available. Instead, we use the F277W band as the green channel. The size of each image is $3^{\prime\prime}\times3^{\prime\prime}$. (b) Medium resolution spectra of the LRDs around the \Ha\ region. The black solid line shows the spectrum along with the errors (gray shaded area). The red solid shows the total single Gaussian fit. The blue dashed line and light blue dotted line show the continuum and the single Gaussian emission, respectively. We also show the corresponding $\chi^2$ value, BIC value, AIC (Akaike Information Criterion, defined as $2k + \chi^2$) value, redshift, and derived intrinsic $\rm FWHM_{H\alpha}$ in each panel.}
 \label{fig:RGB_image_Ha_line}
\end{figure*}

We aim to search LRDs that do not have a broad \Ha\ line component. Among the 98 selected LRDs, 32 have high- or medium-resolution grating spectra covering \Ha, and 20 of them have additional high- or medium-resolution grating spectra covering \Hb\ and $[$O\,\textsc{iii}$]\lambda\lambda~4959,5007$. 
These LRDs with both PRISM and grating spectra are mainly from the RUBIES \citep{deGraaff_2025_RUBIES}, UNCOVER \citep{2024ApJ...974...92B}, and JADES \citep{2023arXiv230602465E} programs.
We fit these lines using \texttt{lmfit} \citep{2021zndo....598352N}. The line spread functions (LSFs) provided by JDOX significantly underestimate the actual spectral resolution \citep[e.g.,][]{2024A&A...684A..87D,2024A&A...691A.145M}, since they apply only for uniformly illuminated sources. Therefore, we use \texttt{msafit} to model the LSF, which forward models the light trajectory on the JWST detector for a given source position to measure the spectral resolution \citep{2024A&A...684A..87D}. Some sources have multiple high- or medium-resolution spectra. In such cases, we use the spectrum with the highest S/N rather than combining them to avoid ambiguity due to the LSF effect. The results from different spectra of the same sources are consistent.

For $[$O\,\textsc{iii}$]\lambda\lambda~4959,5007$, we use a single Gaussian to fit either line separately. We do not apply any constraint  \citep[e.g.,][]{2025ApJ...980...93J}. For \Ha\ and \Hb, we perform two types of fitting:

\begin{enumerate}
    \item A broad Gaussian component (FWHM $> $ 1000 $\rm km~s^{-1}$) and a narrow Gaussian component (FWHM $<$ 1000 $\rm km~s^{-1}$) at the same centroid velocity.
    \item A narrow Gaussian component only (FWHM $<$ 1000 $\rm km~s^{-1}$).
\end{enumerate}
The \NII\ lines are added if needed, and they are forced to have the same width as the narrow component of \Ha. 
We do not tie the FWHM of  \OIII, \Ha, and \Hb\ lines, since we want to check whether they show different line widths later. This choice does not impact the identification of narrow-line LRD candidates in the current step.
To compare the quality of these two types of fitting, we calculate the Bayesian Information Criterion (BIC) parameter, defined as \citep{2007MNRAS.377L..74L}:
\begin{equation}
    {\rm BIC} = \chi^2 + k\ln n,
\end{equation}
where $k$ is the number of free parameters and $n$ is the number of data points. A single Gaussian component is sufficient to describe the \Ha\ line if the following criterion is satisfied:
\begin{equation}
    {\rm BIC}_{\rm H\alpha2} - {\rm BIC}_{\rm H\alpha1} = \Delta {\rm BIC}_{\rm BN} > 1,
\end{equation}
where ${\rm BIC}_{\rm H\alpha2}$ and ${\rm BIC}_{\rm H\alpha1}$ correspond to the two-component fitting and the single-component fitting, respectively.
Based on the above procedure, we identify five narrow-line LRDs with no broad \Ha\ line component from the 32 sources. We also attempt to fit the \Ha\ lines of these five sources using two narrow Gaussian components (FWHM $<$ 1000 $\rm km~s^{-1}$). The resultant BIC value still favors the single narrow component model in each case. The fraction of narrow-line LRDs in our sample is $\sim 16 \%$. This fraction may be biased by the complex selection process of the NIRSpec observations. On the other hand, we argue that the photometric selection is not sensitive to the line width, and thus the narrow-line LRD fraction is unlikely affected significantly.

Some basic properties and the emission line measurement results of these sources are shown in Table \ref{tab:base_info}. The FWHMs have been corrected for the LSF effect that introduces significant uncertainty due to the medium resolution of the spectra, especially for the \OIII\ line. If the difference between the measured FWHM and the intrinsic FWHM is smaller than the propagated uncertainty, this difference is adopted as the upper limit instead. The NIRCam images and the spectral region around \Ha\ for these five LRDs are shown in Figure \ref{fig:RGB_image_Ha_line}. All five LRDs appear very compact and red, resembling typical LRDs. This is expected, as they have passed the rigorous LRD selection criteria. However, their \Ha\ lines are well fitted by a single narrow Gaussian component, mostly with an intrinsic $\rm FWHM_{H \alpha}$ of $150$--$300~\rm km~s^{-1}$. The largest one, UID-16321, has an intrinsic $\rm FWHM_{H \alpha}$ of $475_{-74}^{+42}~\rm km~s^{-1}$. The selected narrow-line LRDs are relatively fainter compared with other LRDs, as shown in Figure \ref{fig:z_mag}. 

\subsection{Morphology}
\label{subsec:morphology_image}

\begin{figure*}
 \includegraphics[width=0.75\textwidth]{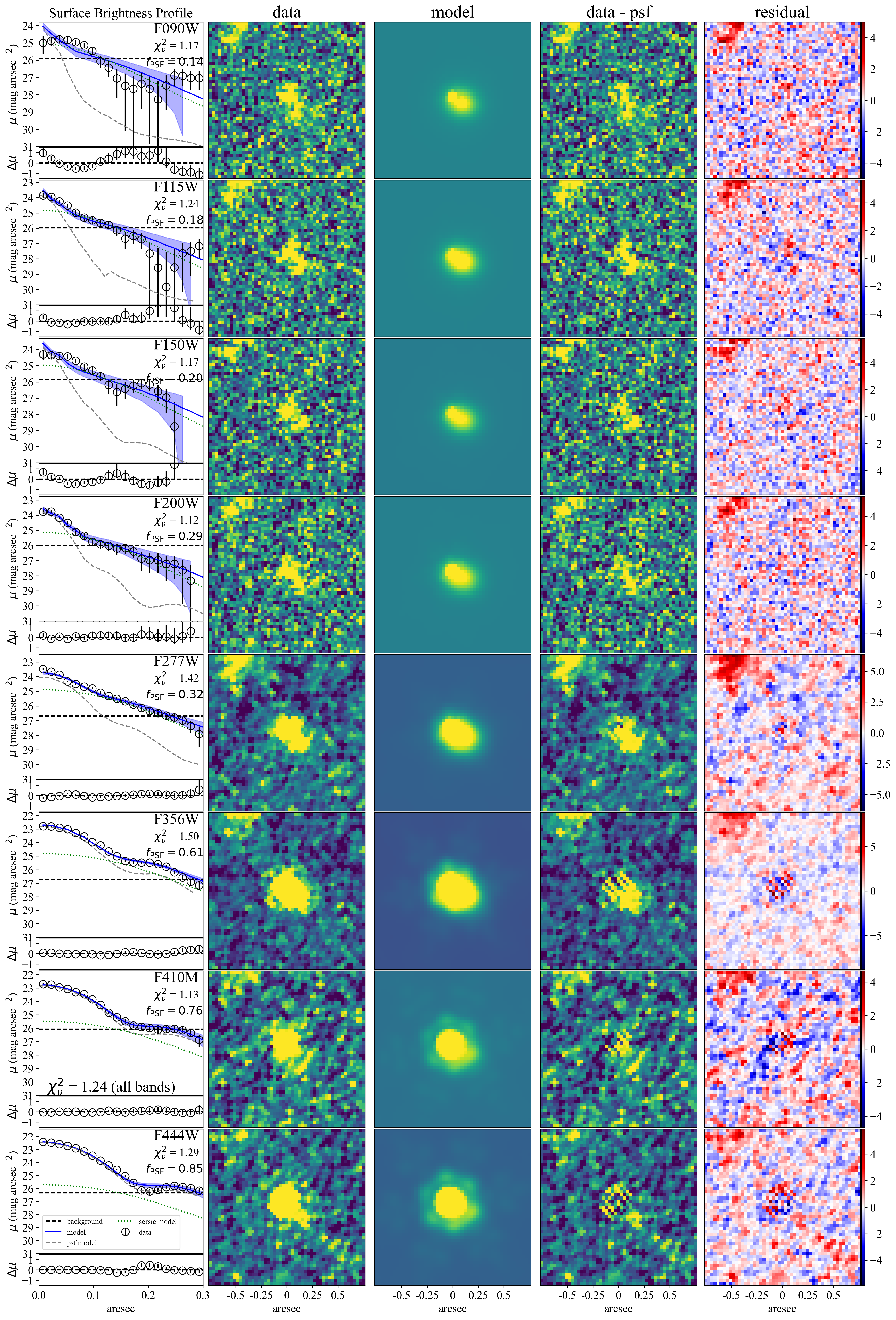}
 \centering
 \caption{Simultaneous multi-band image fitting results using a point-source model and a \sersic\ model for UID-12329. Each row, from top to bottom, shows the results for the eight NIRCam bands (F090W, F115W, F150W, F200W, F277W, F356W, F410M, F444W), respectively. In the left-most column, the upper panel of each row shows the observed radial SB distribution (open circles with error bars), the PSF model (gray dashed line), the \sersic\ model (green dotted line), as well as the total model (blue solid line). The background noise level is denoted by the black horizontal dashed line. The $\chi^2$ value and point-source component fraction for each band are given in the upper-right corner of each panel. The $\chi^2$ value for all eight bands is given in the lower-left corner of the panel for F410M. The lower subpanels give the residuals between the data and the best-fit model (data$-$model). The imaging columns, from left to right, display the original data, best-fit model, data minus the nucleus point-source component, and residuals normalized by the errors (data$-$model$/$error), which are stretched linearly from $-$5 to 5.}
 \label{fig:galfitm_result_12329}
\end{figure*}

The NIRCam images of the five narrow-line LRDs in Figure \ref{fig:RGB_image_Ha_line} show that they are very compact. To quantitatively measure their morphology and detect their potential extended emission, we fit their multi-band JWST images (Figure \ref{fig:galfitm_result_12329}) using \texttt{GalfitM} \citep{2013MNRAS.430..330H,2013MNRAS.435..623V}, a multi-band version of the two-dimensional image fitting code \texttt{Galfit} \citep{2002AJ....124..266P,2010AJ....139.2097P}. We first cut $3^{\prime\prime}\times3^{\prime\prime}$ images for each band from the background-subtracted mosaic images. The corresponding error cutout images are used as input sigma images. To ensure that the best-fit $\chi^2$ value is close to unity, we scale each sigma image by a factor of $\sim 0.5$–$0.75$ such that its median background pixel value matches the standard deviation of the background pixels in the science image. Bands with no significant source detection are excluded during the fitting. For UID-16321, F090W and F115W are excluded; for UID-20547, F090W, F115W, and F150W are excluded; and for UID-8219 and UID-22015, F090W is excluded.

It is still unclear whether LRDs are AGN-dominated or galaxy-dominated, particularly in the case of narrow-line LRDs. Therefore, we perform three types of model fitting: (1) a single PSF model, representing an AGN-dominated morphology; (2) a single \sersic\ model, representing a pure galaxy; and (3) a combination of a PSF and a \sersic\ component, representing an AGN and its host galaxy. The PSF models used are those constructed in Section \ref{subsec:image_data}. We fit a $1\farcs5 \times 1\farcs5$ region centered on the source. During the fitting procedure, the axis ratio, half-light radius, and \sersic\ index of the galaxy component are allowed to vary following a $2^{\rm nd}$-order Chebyshev polynomial, while the position angle is allowed to vary following a $1^{\rm st}$-order Chebyshev polynomial. These flexibilities serve to accommodate the morphological difference between long and short wavelengths, as these sources are often more extended in shorter wavelengths.

For UID-16321, UID-20547, and UID-22015, their multi-band JWST image can be well fitted by a single PSF model. UID-16321 and UID-20547 are very faint in the short wavelengths, so only their long-wavelength data (where LRDs typically appear unresolved) are used to measure the morphology. For UID-20547, it F277W band residual of the PSF fitting suggests weak extended emission. For UID-8219 and UID-12329, their short-wavelength images cannot be well fitted by a single PSF model, revealing clear extended emission. UID-8219 can be sufficiently fitted with a \sersic\ component, and an additional PSF component slightly decreases the $\chi^2$ value from 1.13 to 1.09. In case of UID-12329, a single \sersic\ model also fails to provide a good fit, with significant residuals in the short-wavelength bands. We thus adopt a PSF+\sersic\ model that yields a reasonable result with $\chi^2$ value of 1.24. Figure \ref{fig:galfitm_result_12329} shows the PSF+\sersic\ fitting result for UID-12329. The fraction of the \sersic\ component decreases toward longer wavelengths. The \sersic\ component of UID-12329 is slightly offset from the central point sources; such offset extended emission are also seen in other LRDs \citep[e.g.,][]{2025ApJ...983...60C,2025arXiv250503183C}.

\subsection{Emission-line Properties}
\label{subsubsec:emission_line_prop_BPT}

\begin{figure*}
 \includegraphics[width=0.915\textwidth]{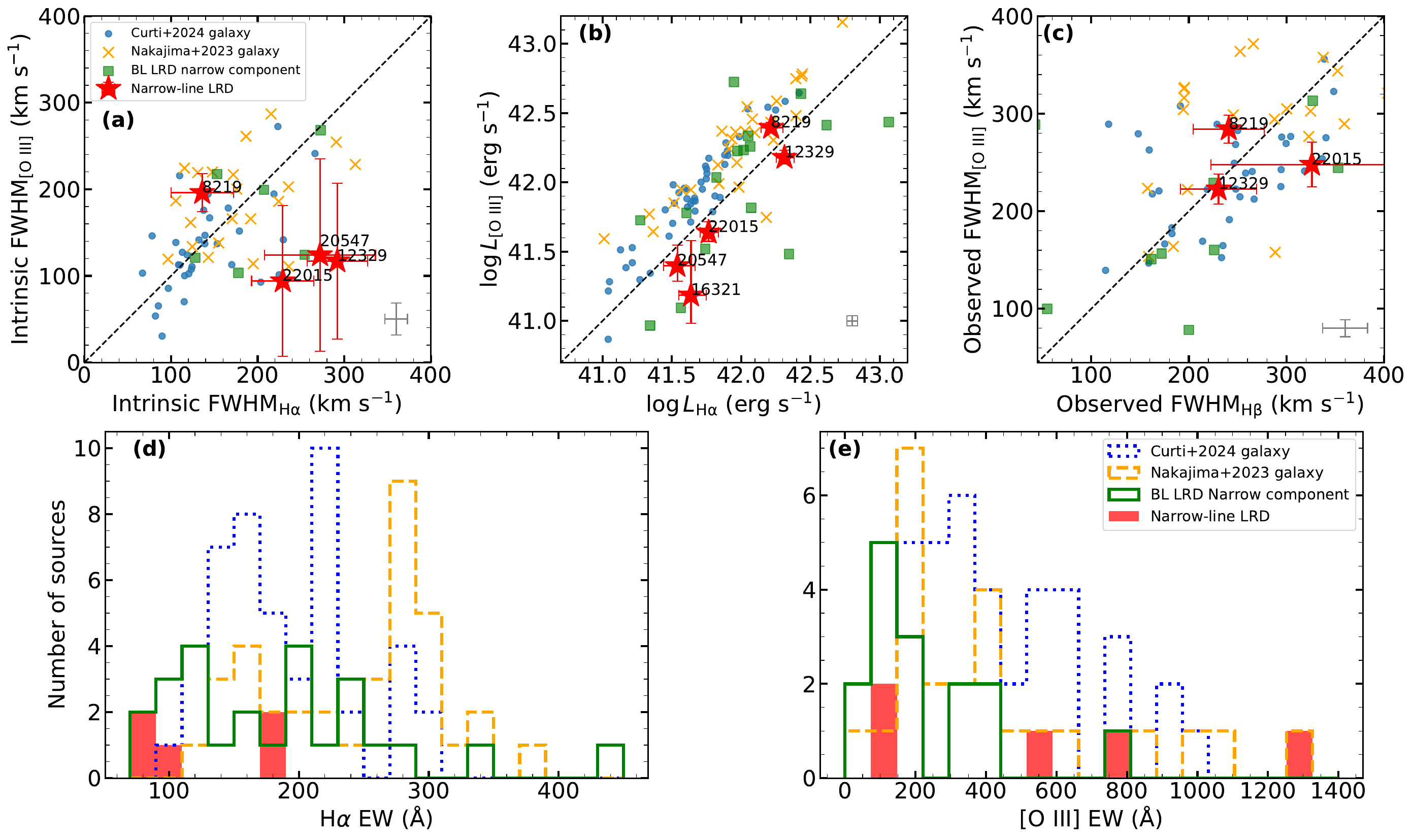}
 \centering
 \caption{Comparison of the intrinsic FWHM in panel (a), and luminosity in panel (b) of \OIII\ and \Ha\ for the five narrow-line LRDs (red stars). Panel (c) shows the comparison of the observed FWHM for \OIII\ and \Hb. We also show star-forming galaxies from \citet[yellow crosses]{2023ApJS..269...33N} and \citet[blue dots]{2024A&A...684A..75C}, and the narrow \Ha/\Hb\ components of broad-line LRDs (green squares) selected in Section \ref{subsec:LRD_selection}. The gray error bars in each panel represent the median measurement errors for the star-forming galaxy sample. Panels (d) and (e) show the EW distributions of the \Ha\ and \OIII\ lines, respectively. }
 \label{fig:Ha_OIII_poperties}
\end{figure*}

Using the emission line quantities measured in Section \ref{subsec:emission_line_fitting}, we analyze the emission line properties of the narrow-line LRDs and compare them with those of normal star-forming galaxies. We adopt galaxy samples from \citet{2023ApJS..269...33N} and \citet{2024A&A...684A..75C} and measure their properties using the same method as applied to our sources. Figure \ref{fig:Ha_OIII_poperties} presents a comparison of the intrinsic FWHM, luminosity, and equivalent width (EW) of the \OIII, \Ha, and \Hb\ lines among narrow-line LRDs, broad-line LRDs, and the star-forming galaxies. For broad-line LRDs, we only use the narrow components of the \Ha\ and \Hb\ lines for comparison. UID-16321 does not have medium- or high-resolution spectral coverage of the \OIII\ line, thus its $\rm FWHM_{[O,\textsc{iii}]}$ values are not available. We measure its $L_{[\rm O,\textsc{iii}]}$ from the prism spectrum. The comparison of the FWHM between the \OIII\ and \Hb\ lines in Figure \ref{fig:Ha_OIII_poperties}(c) does not account for the LSF effect, as these two lines have similar wavelengths and spectral resolutions.

Figure \ref{fig:Ha_OIII_poperties} suggests that narrow-line LRDs may exhibit different emission-line properties compared to typical star-forming galaxies. 
NIRSpec grating spectrum typicall has higher spectral resolutions at longer wavelengths. 
After correcting the LSF effect, $\rm FWHM_{H\alpha}$ is consistent with $\rm FWHM_{[O\,\textsc{iii}]}$ (Figure \ref{fig:Ha_OIII_poperties}a). For the narrow-line LRDs, three of them have $\rm FWHM_{H\alpha}$ slightly larger than $\rm FWHM_{[O\,\textsc{iii}]}$. On the other hand, three LRDs have medium-resolution spectra covering \Hb\ and they all show consistent $\rm FWHM_{H\beta}$ and $\rm FWHM_{[O\,\textsc{iii}]}$. UID-22015 shows tentative larger $\rm FWHM_{H\beta}$ but its measurement error is $\sim 104 \rm ~km~s^{-1}$. 
The observed $L_{\rm H\alpha}$ of narrow-line LRDs also tend to be larger compared with the star-forming galaxies with similar $L_{[\rm O\,\textsc{iii}]}$ as shown in Figure \ref{fig:Ha_OIII_poperties}(b). 
We note the line luminosities here are not corrected for dust attenuation. The larger \Ha\ luminosities of narrow-line LRDs may simply reflect higher dust extinction, as \OIII\ is more strongly attenuated than \Ha. However, only mild dust extinction is expected for LRDs \citep[e.g.,][]{2025arXiv250518873C,2025arXiv250522600C,2025arXiv250302059S}.
Conversely, the narrow-line LRDs generally exhibit smaller \Ha\ EWs and larger \OIII\ EWs compared with star-forming galaxies, although two narrow-line LRDs have relatively low \OIII\ EWs. These different emission-line properties, along with their characteristic V-shape SEDs, suggest that narrow-line LRDs represent a special population. 

\begin{figure}[t]
 \includegraphics[width=0.46\textwidth]{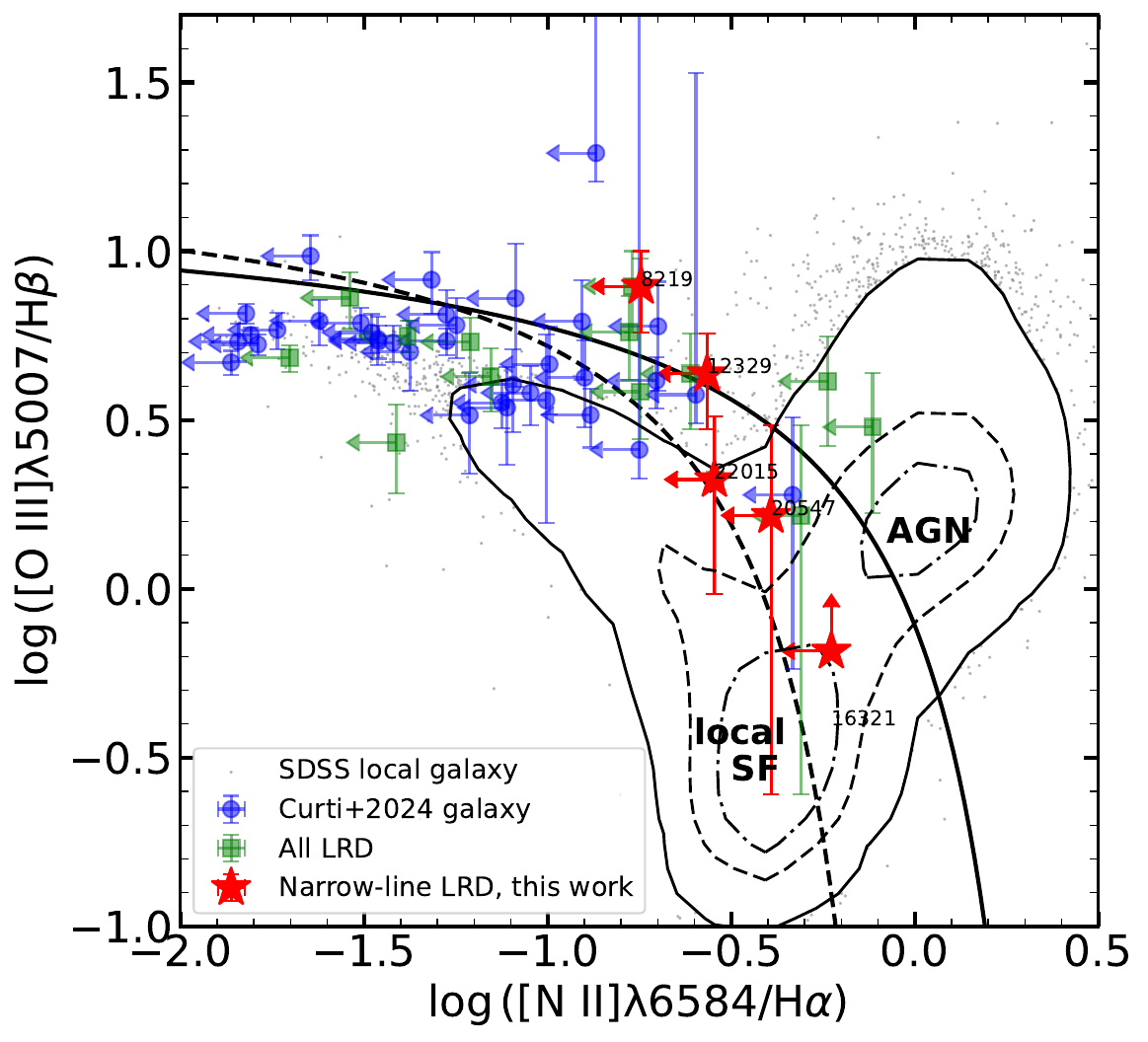}
 \centering
 \caption{BPT diagram. The five narrow-line LRDs are shown as the red stars and other LRDs are shown as the green squares. The contours represent objects at $0<z<0.5$ from SDSS \citep[][encloses 35, 68, and 95 percent of the objects, respectively]{2023ApJS..267...44A}, showing the AGN branch and the star-forming galaxy branch. The star-forming galaxies in \citet{2024A&A...684A..75C} are also shown as the blue dots for comparison.}
 \label{fig:BPT_diagram}
\end{figure}

We also check the locations of the narrow-line LRDs and other LRDs in the Baldwin, Phillips, \& Telervich (BPT) diagnostic diagrams \citep[][]{1981PASP...93....5B,1987ApJS...63..295V}, using the measured narrow component properties. For all LRDs, the \NII\ doublets are very faint and undetected, allowing only upper limits for the \NII/\Ha\ ratios (a 3$\sigma$ upper limit is adopted). For UID-16321, its \Hb\ line is not detected in both the prism and grating spectrum. We put an upper limit on its \OIII/\Hb\ ratio based on the measured 3$\sigma$ flux upper limit of the \Hb\ line in the grating spectrum. The resulting distribution of the five narrow-line LRDs on the BPT diagram is shown in Figure \ref{fig:BPT_diagram}, in which the SDSS low-redshift galaxies and JWST high-redshift star-forming galaxies are plotted for comparison. The solid and dashed lines indicate the demarcation between local AGN and star-forming galaxies from \citet{2001ApJ...556..121K} and \citet{2003MNRAS.346.1055K}, respectively. The narrow-line LRDs, as well as other LRDs, lie near the demarcation. Additionally, the traditional BPT diagram is likely not applicable to high-redshift objects, as high-redshift AGNs are offset from the local AGN branch and mix with the star-forming galaxy branch \citep[e.g.,][]{2023ApJ...954L...4K,2023ApJ...959...39H,2025arXiv250403551J}. Therefore, we are not able to distinguish AGNs from star-forming galaxies based on this diagram.

\subsection{Composite Spectra}

\begin{figure*}
 \includegraphics[width=0.9\textwidth]{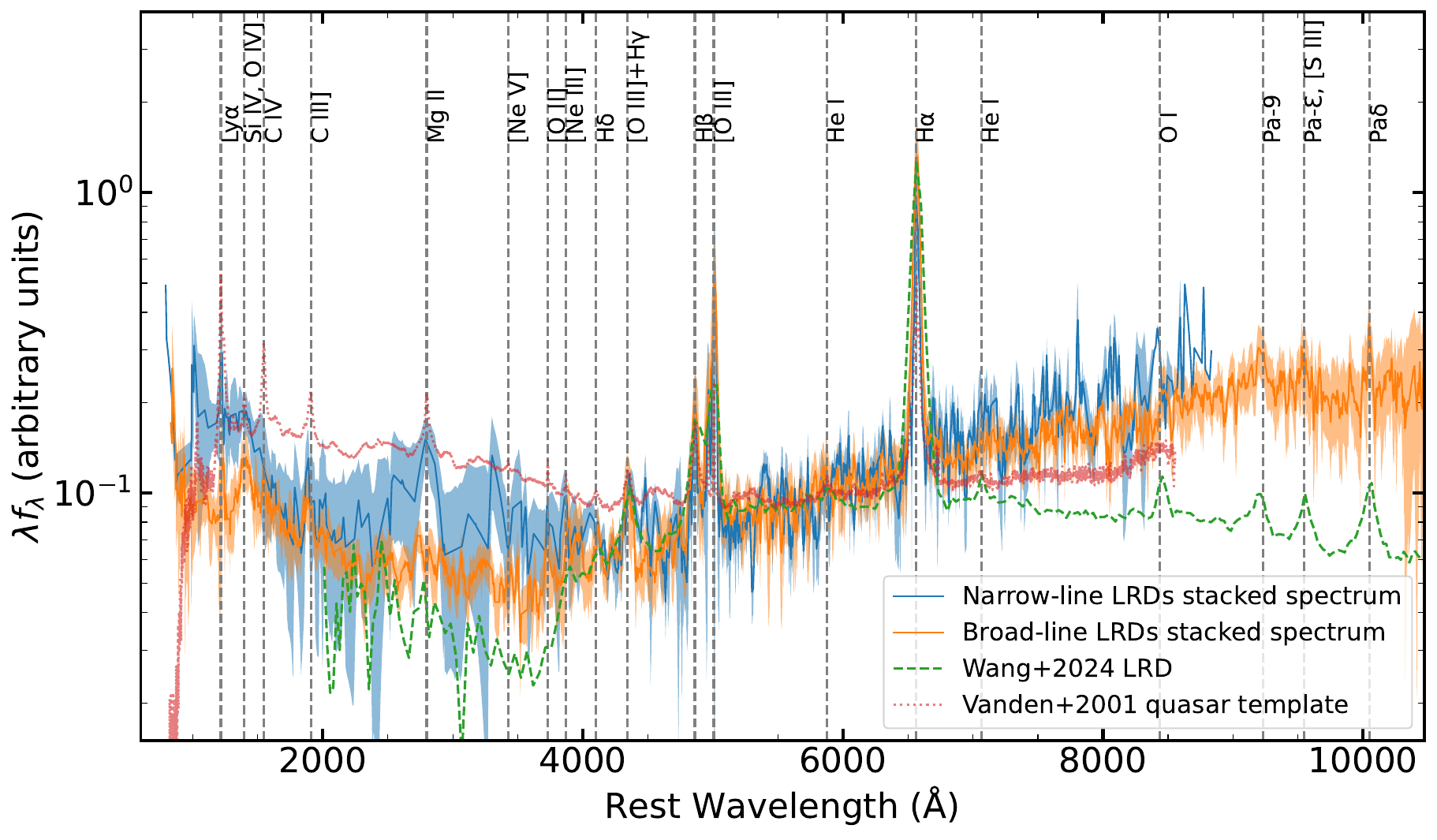}
 \centering
 \caption{Median-stacked spectra of narrow-line LRDs (blue solid curve) and broad-line LRDs (orange solid curve). For comparison, we also plot the LRDs at $z = 3.1$ from \citet[green dashed curve]{2024arXiv240302304W} and the composite spectrum of typical SDSS quasars \citep[red dotted curve]{2001AJ....122..549V}.  The spectra are normalized at 5000\AA. Main emission lines are marked.}
 \label{fig:stacked_spec}
\end{figure*}

We produce composite spectra for narrow-line LRDs and broad-line LRDs using their NIRSpec/PRISM spectra, and compare the composite spectra in Figure \ref{fig:stacked_spec}. The composite spectra are generated using median stacking without normalization. Mean stacking yields consistent results. The uncertainty is estimated by measuring the standard deviation from 1000 bootstrap realizations of the sample. The prism spectra of the five narrow-line LRDs generally have low S/Ns due to their faintness, resulting in large errors in the composite spectrum. UID-12329 only has coverage in the rest-frame 4100--7500 $\AA$ range because it lies near the detector gap. Overall, the continuum slopes and emission lines of the stacked narrow-line and broad-line LRD spectra are very similar, except that the narrow-line LRDs show a narrower \Ha\ line, confirming our selection.

\section{Interpretation of Narrow-line LRDs}
\label{sec:two_scenario}

In this section, we consider two possible interpretations of the narrow-line LRDs: they may be either compact star-forming galaxies or AGNs with low black hole masses. There is another possibility that narrow-line LRDs represent the ``Type 2'' counterparts of the broad-line LRDs, with the difference arising from obscuration by a dust torus. However, given the similar SEDs of the broad-line and narrow-line LRDs, this scenario appears puzzling and is not considered further in this section.  We will briefly revisit it in Section \ref{subsec:implication}.

\subsection{Purely Star-forming Galaxies}
\label{subsec:pure_SF_gala}

The presence of broad Balmer lines in LRDs is the key evidence of AGN activity. Conversely, the absence of a broad H$\alpha$ component suggests that the AGN contribution in these narrow-line LRDs is likely negligible. The extended UV emission detected in some of the narrow-line LRDs (see Section \ref{subsec:morphology_image}) also supports significant galaxy contribution. Therefore, narrow-line LRDs can be purely star-forming galaxies with a special V-shape continuum.

\subsubsection{Stellar Masses and Star Formation Rates}
\label{subsubsec:star_mass_SFR_SEDmodeling}

To estimate their physical properties, including stellar mass and star formation rate (SFR), we perform SED modeling for the narrow-line LRDs using \texttt{CIGALE} and \citep[v2025,][]{2005MNRAS.360.1413B,2019A&A...622A.103B} and \texttt{Bagpipe} \citep{2018MNRAS.480.4379C,2019MNRAS.490..417C}, assuming that the whole SED is produced by galaxy. We employ two different codes with distinct configurations to explore the redundancy of the best-fit SED solutions. Each source's redshift is fixed at the spectroscopic redshift provided by DJA. 

For \texttt{CIGALE}, we assume a delayed-$\tau$ star formation history (SFH), where ${\rm SFR} \propto te^{-t/\tau}$ and $\tau \in [0.01, 2]\rm \,Gyr$. We also allow an optional late starburst in the latest 20$\, \rm Myr$. We use the \citet{2003MNRAS.344.1000B} stellar population synthesis models and adopt the \citet{2003PASP..115..763C} initial mass function (IMF). Nebular continuum and emission lines are included, allowing a metallicity range of [0.2, 1]$\,Z_\odot$ and an ionization parameter $\log U\in[-4, -1]$. We adopt the SMC extinction curve \citep{1992ApJ...395..130P} for nebular emission and the modified \citet{2000ApJ...533..682C} attenuation curve for the stellar continuum, with the power-law slope $n \in [-0.6, 0.2]$. We add a relative error of 5\% in quadrature to the uncertainties of the fluxes.

\begin{figure*}
 \includegraphics[width=0.98\textwidth]{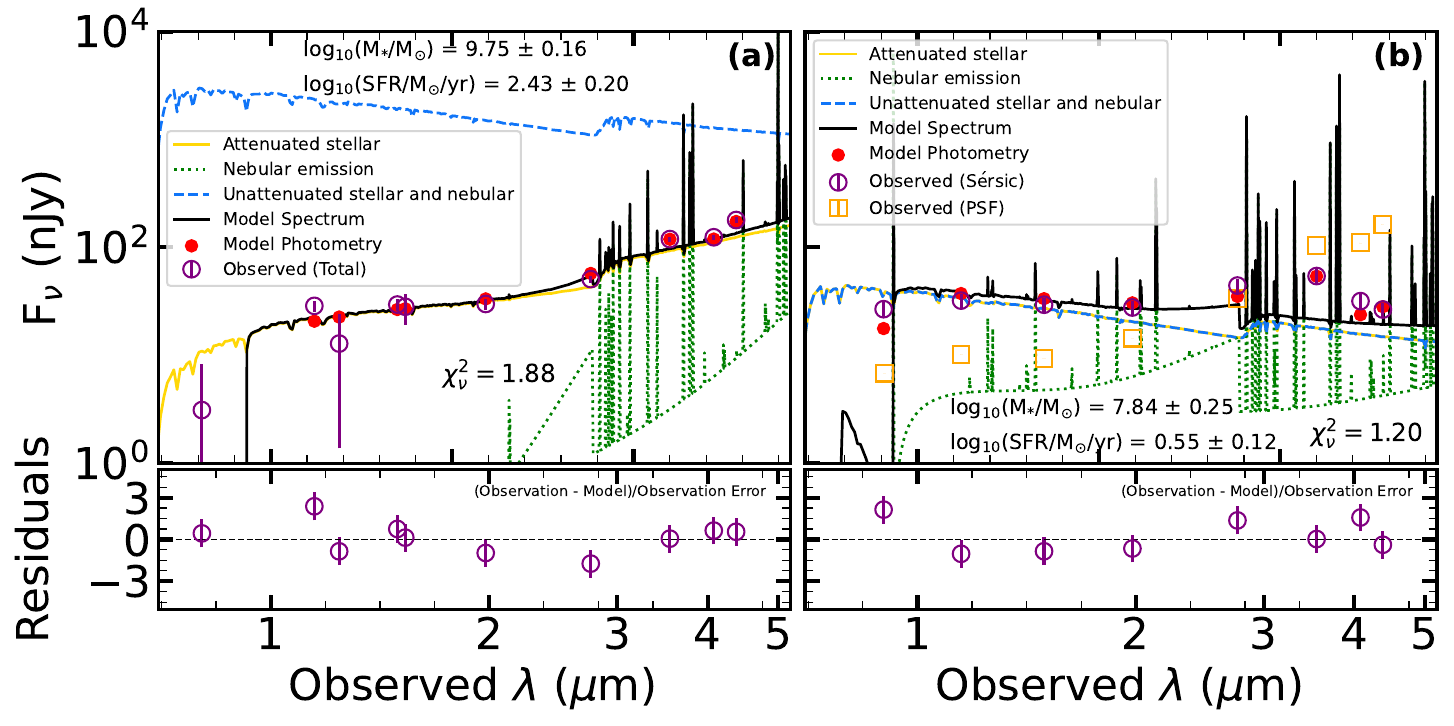}
 \centering
 \caption{Best-fit SED models of UID-12329 by \texttt{CIGALE} using a pure galaxy model described in Section \ref{subsubsec:star_mass_SFR_SEDmodeling}, assuming (a) all emission originates from the galaxy and (b) only the extended emission originates from the galaxy. The bottom panels show the relative residuals between the observed and modeled photometry. In panel (b), the fitting is performed only for the observed \sersic\ component, and the SED of the decomposed PSF component is shown as the orange squares for comparison. The derived stellar mass and SFR are shown in each panel. }
 \label{fig:cigale_12329}
\end{figure*}

For \texttt{Bagpipe}, we use the synthetic templates from \citet{2003MNRAS.344.1000B} with a Kroupa IMF, a stellar mass cut-off of $100~M_{\odot}$, and nebular emission computed self-consistently with \texttt{CLOUDY} \citep{2013RMxAA..49..137F} with $\log U\in[-4, -1]$. We assume a delayed-$\tau$ SFH, where ${\rm SFR} \propto te^{-t/\tau}$ and $\tau \in [0.01, 2]\rm \,Gyr$. We also allow an optional late starburst in the latest 20$\, \rm Myr$. The allowed metallicity range is [0.2, 1]$\,Z_\odot$. We adopt the double-component attenuation law of \citet{2000ApJ...539..718C} with the power-law slope $n \in [0.3, 2.5]$. The low-resolution prism spectrum and the multi-band photometric data are fitted simultaneously in the \texttt{Bagpipe} fitting. We adopt a flexible calibration scheme to account for potential discrepancies in relative flux calibration between spectra and photometry. Specifically, we model the spectroscopic flux calibration uncertainties by fitting a third-order Chebyshev polynomial to the spectrum during the fitting process. This allows us to capture effects such as aperture mismatch and wavelength-dependent calibration issues. To deal with potentially underestimated errors, we also add a noise model component to fit a multiplicative factor to all of the spectroscopic uncertainties.

Both \texttt{CIGALE} and \texttt{Bagpipes} yield reasonable best-fit SED models. As an example, the \texttt{CIGALE} best-fit result for UID-12329 is shown in Figure~\ref{fig:cigale_12329}(a). The SED modeling result suggests that the broad-band SED of a narrow-line LRD can be explained as a galaxy with a high stellar mass and SFR. It shows strong dust attenuation, and the characteristic V-shape is caused by the Balmer break and differential attenuation at rest-frame UV and optical wavelengths. This is consistent with the initial interpretation of LRDs \citep[e.g.,][]{2023Natur.616..266L,2023ApJ...956...61A,2024ApJ...968....4P}, which is expected since the broad-band SED properties of broad-line and narrow-line LRDs are similar.

\begin{figure}
\hspace{-0.4cm}
\includegraphics[width=0.49\textwidth]{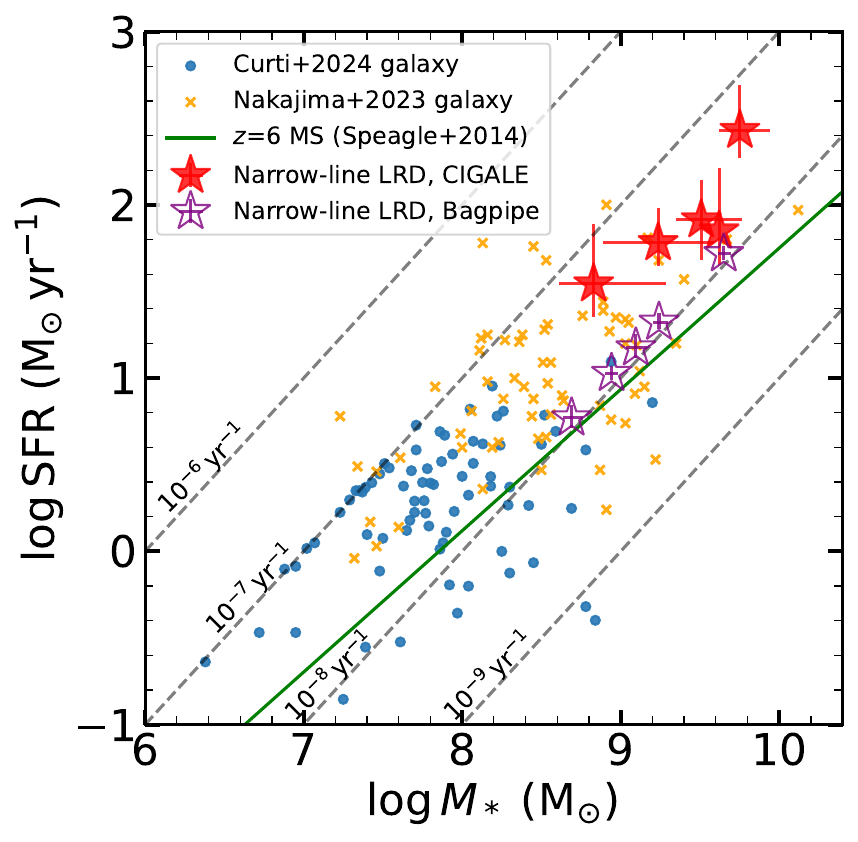}
 \centering
 \caption{Stellar masses and SFRs of the narrow-line LRDs estimated by \texttt{CIGALE} (red filled stars) and \texttt{Bagpipe} (purple open stars). They are compared with star-forming galaxies from \citet{2023ApJS..269...33N} and \citet{2024A&A...684A..75C}. The symbols used are the same as those in Figure \ref{fig:Ha_OIII_poperties}. We also show the best-fit main sequence at $z = 6$ in \citet{2014ApJS..214...15S} for comparison. The background lines represent sSFR values ranging from $10^{-9}$ to $10^{-6}~ \rm yr^{\rm -1}$, from bottom to top.}
 \label{fig:Mstar_SFR}
\end{figure}

The inferred stellar masses and SFRs of the narrow-line LRDs are listed in Table \ref{tab:stellar_BH_property} and plotted in Figure \ref{fig:Mstar_SFR}. Star-forming galaxies from \citet{2023ApJS..269...33N} and \citet{2024A&A...684A..75C}, and broad-line LRDs are also shown for comparison. The narrow-line LRDs are located at the large stellar mass and SFR end compared to the normal star-forming galaxies. The stellar masses and SFRs inferred from the \texttt{CIGALE} and \texttt{Bagpipe} show deviations of $\sim 0.5$ dex, with the \texttt{CIGALE} results generally have larger stellar masses and SFRs. Such differences are commonly seen in galaxy-only SED modeling of LRDs \citep[e.g.,][]{2023ApJ...956...61A}. The discrepancy is possibly driven by different prescriptions for the SFH, dust attenuation law, IMF, and the inclusion of spectrum information in the \texttt{Bagpipe} fitting. Nonetheless, assuming either of the two measurement values does not alter any of the conclusions presented in this work. 

\begin{table*}
\setlength{\tabcolsep}{2pt}
\caption{Stellar and Black Hole Properties of the Five Narrow-line LRDs}
\label{tab:stellar_BH_property}
\hspace*{-2.85cm}
\begin{tabular}{ccccccccccccc} 
\hline
\hline
UID & $\lg M_{*,\rm cig}$ & $\lg M_{*,\rm bag}$ & $\rm 
 \lg SFR_{\rm cig}$ & $\rm 
 \lg SFR_{\rm bag}$ & $\rm \lg SFR_{\rm NUV}$ & $\rm \lg SFR_{\rm H\alpha}$ & $\lg M_{*,\rm Sersic}$ & $\lg M_{\rm BH}$ & $\lg \sigma_{\rm B18}$ & $\lg \sigma_{\rm H09}$ & $\lg M_{\rm dyn,B18}$ & $\lg M_{\rm dyn,H09}$    \\ 
 & $\rm M_\odot$ & $\rm M_\odot$ & $\rm M_\odot ~\rm yr^{-1}$ & $\rm M_\odot ~\rm yr^{-1}$ & $\rm M_\odot ~\rm yr^{-1}$ & $\rm M_\odot ~\rm yr^{-1}$ & $\rm M_\odot$ & $\rm M_\odot$ & $\rm km~s^{-1}$  & $\rm km~s^{-1}$ &  $\rm M_\odot$ &  $\rm M_\odot$ \\
(1)&(2)&(3)&(4)&(5)&(6)&(7)&(8)&(9)&(10)&(11)&(12)&(13)\\
\hline
8219 & $9.51^{+0.16}_{-0.26}$ & $8.94^{+0.04}_{-0.04}$ & $1.91^{+0.23}_{-0.23}$ & $1.03^{+0.04}_{-0.04}$ & 1.14 & 1.58 & $7.42^{+0.16}_{-0.26}$ & $<5.9$ & $2.02^{+0.06}_{-0.06}$ & $1.83^{+0.06}_{-0.06}$ & $9.27^{+0.50}_{-0.50}$ & $8.90^{+0.44}_{-0.44}$ \\ 
\hline
12329 & $9.75^{+0.13}_{-0.19}$ & $9.65^{+0.04}_{-0.05}$ & $2.43^{+0.16}_{-0.26}$ & $1.72^{+0.04}_{-0.04}$ & 2.02 & 1.92 & $7.84^{+0.25}_{-0.25}$ & $<6.0$ & $1.90^{+0.19}_{-0.34}$ & $1.59^{+0.19}_{-0.34}$ & $9.23^{+0.47}_{-0.47}$ & $8.62^{+0.47}_{-0.47}$\\ 
\hline
16321 & $9.24^{+0.35}_{-0.35}$ & $9.10^{+0.06}_{-0.07}$ & $1.78^{+0.14}_{-0.20}$ & $1.18^{+0.06}_{-0.08}$ & 1.02 & 1.42 & -- & $<5.8$ & -- & -- & -- & --\\ 
\hline
20547 & $9.62^{+0.36}_{-0.21}$ & $9.24^{+0.04}_{-0.06}$ & $1.85^{+0.20}_{-0.37}$ & $1.32^{+0.04}_{-0.05}$ & 1.89 & 1.27 & -- & $<5.6$ &$1.92^{+0.19}_{-0.36}$ & $1.65^{+0.19}_{-0.36}$ & $9.23^{+0.49}_{-0.49}$ & $8.69^{+0.49}_{-0.49}$\\ 
\hline
22015 & $8.83^{+0.22}_{-0.46}$ & $8.69^{+0.06}_{-0.07}$ & $1.54^{+0.19}_{-0.35}$ & $0.77^{+0.06}_{-0.07}$  & 1.14 & 1.17 & -- & $<5.5$ & $1.85^{+0.18}_{-0.32}$ & $1.53^{+0.18}_{-0.32}$ & $9.37^{+0.45}_{-0.45}$ & $8.72^{+0.45}_{-0.45}$\\
\hline
\hline
\end{tabular}
\tablecomments{
Col. (1): UID in the DJA catalog.
Cols. (2) and (3): Stellar masses of the best-fit \texttt{CIGALE} and \texttt{Bagpipe} models.
Cols. (4) and (5): SFRs of the best-fit \texttt{CIGALE} and \texttt{Bagpipe} models.
Cols. (6) and (7): SFRs derived using NUV luminosity and \Ha\ luminosity.
Col. (8): Stellar mass of the extended emission derived using \texttt{CIGALE}. 
Col. (9): Black hole mass.
Cols. (10) and (11): Gas velocity dispersion using the \citet{2018ApJ...868L..36B} and \citet{2009ApJ...699..638H} corrections.
Cols. (12) and (13): Dynamical mass using the \citet{2018ApJ...868L..36B} and \citet{2009ApJ...699..638H} corrections.
}
\end{table*}

\subsubsection{Comparison of UV-based and H$\alpha$-based SFRs}
\label{subsec:UV_Ha_SFR_comparision}

\begin{figure*}
 \includegraphics[width=0.95\textwidth]{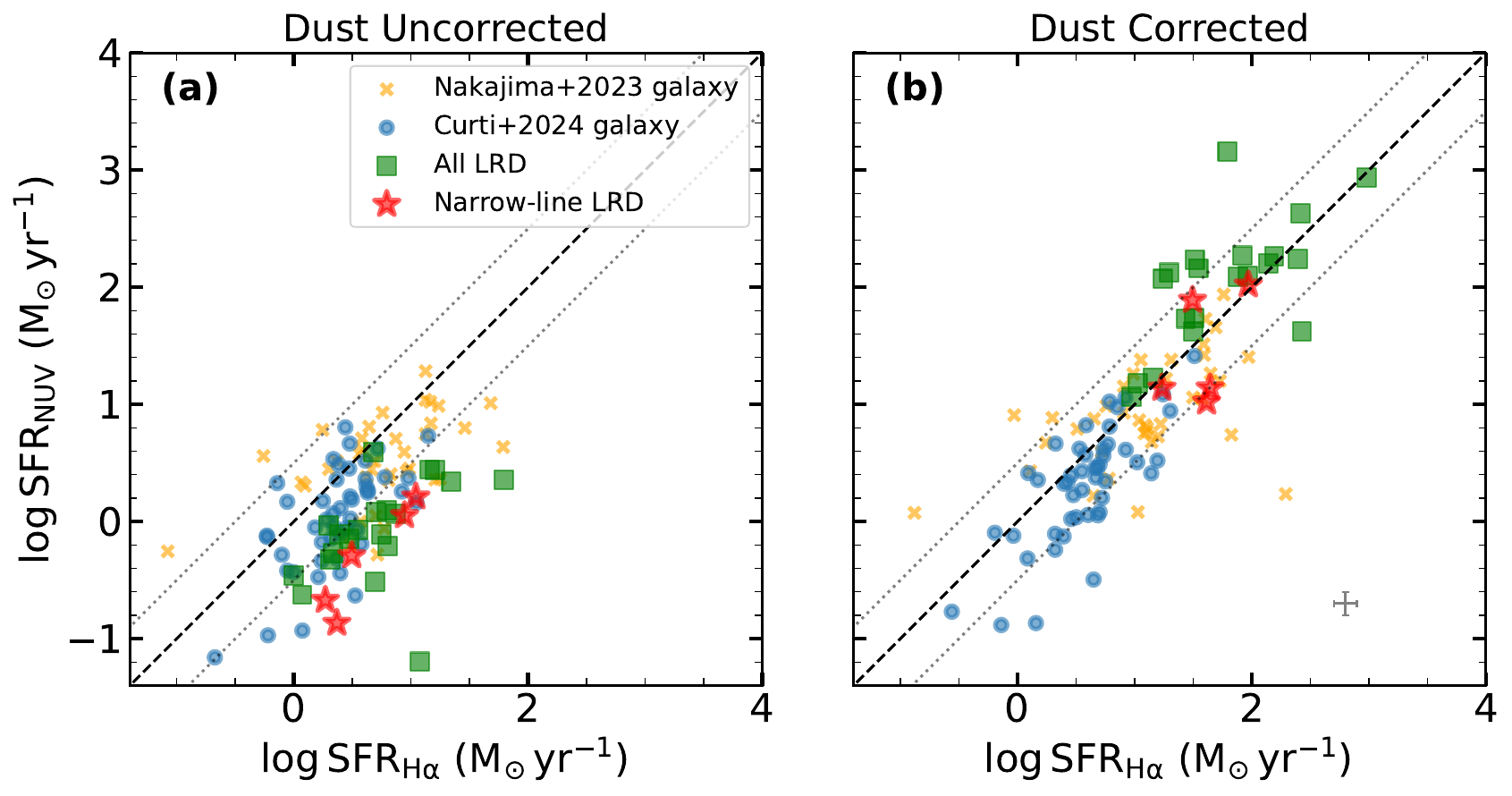}
 \centering
 \caption{Comparison of UV-based and H$\alpha$-based SFRs with (a) and without (b) dust correction for LRDs. We also show star-forming galaxies from \citet{2023ApJS..269...33N} and \citet{2024A&A...684A..75C} for comparison, with $L_{\rm NUV}$ and $L_{\rm H\alpha}$ measured using the same method as that applied to LRDs. The symbols used are the same as those in Figure \ref{fig:Ha_OIII_poperties}. The dashed line represents a 1:1 relation. The dotted lines represent a deviation of 0.5 dex. }
 \label{fig:UV_Ha_SFR_compare}
\end{figure*}

In this subsection, we assume that the whole SEDs of the LRDs are from galaxies and compare the SFRs derived from the UV continuum emission and $\rm H\alpha$ line emission. The UV emission directly traces the photospheric emission of young stars, tracing recent star formation over the past 10--200 Myr. The \Ha\ line emission traces ionized gas surrounding massive young stars, providing a nearly instantaneous measure of the SFR by probing stars with lifetimes of 3--10 Myr. They are two of the most commonly used tracers of SFR and are both sensitive to the recent star formation. We adopt  $L_{\rm H\alpha}$ measured in Section \ref{subsec:emission_line_fitting} and  $L_{\rm NUV}$ at rest-frame 1550 $\AA$ from the best-fit SED model obtained using \texttt{CIGALE}, as described in Section \ref{subsubsec:star_mass_SFR_SEDmodeling}. For broad-line LRDs, we use the narrow-line $\rm H\alpha$ component from the two-Gaussian fitting to calculate  $\rm SFR_{H\alpha}$. The SFRs are derived using the calibrations from \citet{2011ApJ...741..124H} and \citet{2011ApJ...737...67M}, as summarized in \citet{2012ARA&A..50..531K}:

\begin{equation}
    \log {\rm SFR_{\rm NUV} ~(M_{\odot}~yr^{-1})} = \log L_{\rm NUV} - 43.17   
\end{equation}
and
\begin{equation}
   \log {\rm SFR_{\rm H\alpha} ~(M_{\odot}~yr^{-1})} = \log L_{\rm  H\alpha} - 41.27.
\end{equation}

We report the $\rm SFR_{\rm NUV}$ and $\rm SFR_{\rm H\alpha}$ values in Table \ref{tab:stellar_BH_property}. 
In our analysis here, the dust attenuation is derived from the CIGALE SED fitting in Section \ref{subsubsec:star_mass_SFR_SEDmodeling}. For each source, we use the best-fit color excess $E(B-V)_{\rm line}$ for nebular lines and convert it to $A_{\rm H\alpha} = 3.16 E(B-V)_{\rm line}$. Dust-corrected \Ha\ luminosities are then calculated using $L_{\rm H\alpha,corr} = L_{\rm H\alpha,obs}\times10^{0.4 A_{\rm H\alpha}} $.
The UV dust correction is computed by the model SEDs provided by CIGALE: we evaluate the model flux density at the rest-frame NUV pivot wavelengths before and after the attenuation correction to obtain dust-corrected UV luminosities.
The comparisons of the UV-based and H$\alpha$-based SFRs with and without dust correction for LRDs are shown in Figure \ref{fig:UV_Ha_SFR_compare}, along with those of the star-forming galaxies from \citet{2023ApJS..269...33N} and \citet{2024A&A...684A..75C} measured by the same method. The \sfrha\ and \sfruv\ are consistent for star-forming galaxies in both the dust-uncorrected and dust-corrected cases, indicating modest and similar levels of dust extinction affecting their UV and \Ha\ emission. In contrast, for LRDs, including narrow-line LRDs, \sfrha\ is significantly higher than \sfruv\ when no dust correction is applied, but the two become consistent after the dust correction. This result suggests that UV continuum and narrow \Ha\ emission in LRDs are likely dominated by star formation, which supports the purely star-forming galaxy scenario, whereas the broad $\rm H\alpha$ component may originate from a non-AGN mechanism, such as Raman scattering \citep[see discussion in ][for possible origins]{2024arXiv240704777K, 2025MNRAS.538.1921M}.

\sfrha\ and \sfruv\ can be used to calculate the burstiness parameter $B = \log(\rm SFR_{\rm H\alpha}/SFR_{\rm UV})$ that is an indicator of the current burst state of a galaxy \citep{2019ApJ...873...74B}. The average $B$ parameter of the narrow-line LRDs is nearly 0 if use the dust-corrected results, implying that the galaxy population does not have a rapidly rising or falling star formation recently. The scatters of the $B$ parameter are similar for narrow-line LRDs and normal star-forming galaxies, indicating their similar bursty levels.

In this purely star-forming galaxies scenario, the inferred SFRs are quite high (up to $1000 ~\rm M_{\odot}~yr^{-1}$, see Figure \ref{fig:UV_Ha_SFR_compare}) for broad-line LRDs, which is contradicted to the lack of strong IR and \CII\ emission for most LRDs, as revealed by deep ALMA/NOEMA observations \citep[][]{2023Natur.616..266L,2024ApJ...968...34W,2025arXiv250522600C,2025arXiv250518873C,2025arXiv250301945X,2025arXiv250302059S}. These observations have primarily targeted relatively bright LRDs with broad-line detections, so it remains unclear whether the same trend applies to our narrow-line LRDs. Moreover, narrow-line LRDs tend to have more moderate SFRs (typically $\sim 100~\rm M_{\odot}~yr^{-1}$), and their expected IR emission is correspondingly weaker.

\subsubsection{Balmer Decrement}

\begin{figure}
 \includegraphics[width=0.45\textwidth]{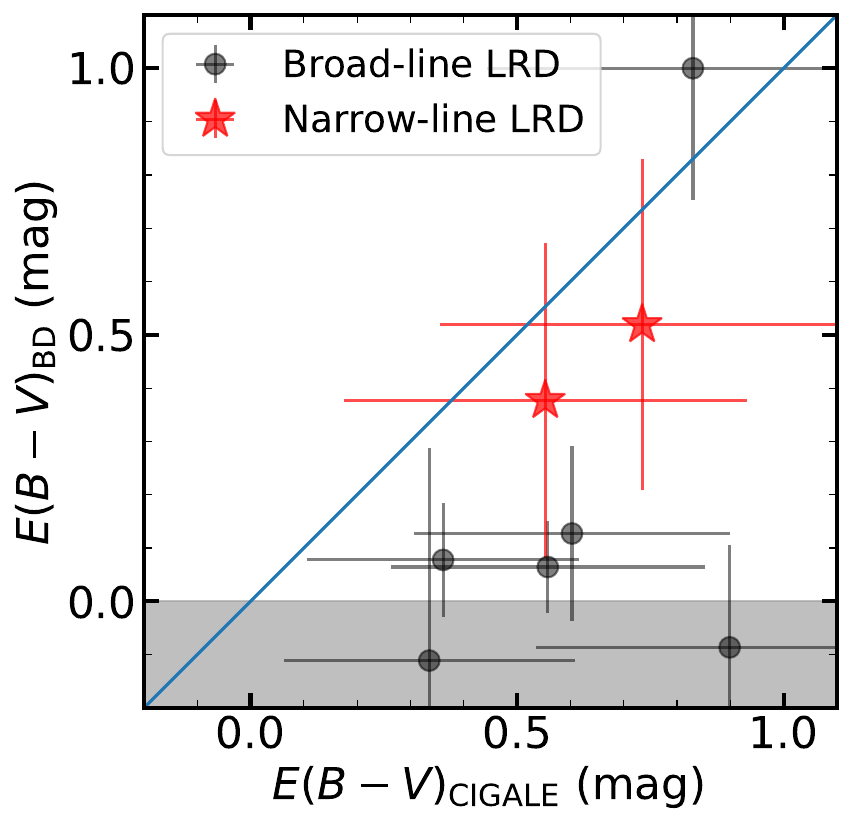}
 \centering
 \caption{Comparison between $E(B-V)$ derived from the SED fitting of \texttt{CIGALE} ($E(B-V)_{\rm CIGALE}$) and from the extinction based on the Balmer decrement ($E(B-V)_{\rm BD}$). Red stars and gray dots represents Narrow-line LRDs and Broad-line LRDs, respectively. The solid blue line represent 1:1 relation. The gray shaded area is where the objects have $E(B-V)_{\rm BD} < 0$ due to their \Ha /\Hb\ line ratio being lower than the canonical Case B value of 2.86.}
 \label{fig:EBV_comparision}
\end{figure}

In our sample, there are eight LRDs (including two narrow-line LRD UID-8219 and UID-12329) with \Hb\ line detections in the high- or medium-resolution spectra with $\rm S/N > 3$. In the galaxy-only scenario, we further check the consistency of the stellar extinction derived from the SED fitting and the nebular extinction from the line flux ratio of \Ha\ and \Hb, i.e., the Balmer decrement. For broad-line LRDs, we only use their narrow-line components. The detailed calculation can be found in Appendix \ref{appendix:balmer_decrement}. The comparison of the two $E(B - V)$ measurements is shown in Figure \ref{fig:EBV_comparision}. For the two narrow-line LRDs, their $E(B - V)_{\rm BD}$ and $E(B - V)_{\rm CIGALE}$ values are roughly consistent, suggesting that the galaxy-only SED fitting results are reasonable. In contract, broad-line LRDs tend to have larger $E(B - V)_{\rm CIGALE}$, indicating that the galaxy-only SED model is not sufficient to describe these sources. The $E(B - V)_{\rm BD}$ values of two LRDs are even smaller than zero. This is because they have \Ha/\Hb\ line ratios smaller than 2.86, which is different from the Case B assumption. Similar results have been reported by previous works \citep[e.g.,][]{2024ApJ...969...90P,2025arXiv250321896S}. 

\subsection{AGNs with Low Black Hole Masses}
\label{subsec:low_mass_BH}

A large fraction of LRDs are believed to be AGNs, as indicated by the presence of broad Balmer lines. The observed V-shape continuum may result from AGN emission attenuated by very dense, dust-free gas, possibly combined with a scattered or leaked AGN component or host galaxy light \citep[e.g.,][]{2024ApJ...964...39G,2025ApJ...980L..27I,2025arXiv250113082J,Lin2025_locallrd,LiuH_2025,2025arXiv250316596N,Zhang2025_lensedlrd}.  
Alternatively, the V-shape may result from a UV-flattened dust-extinction law \citep[e.g.,][]{2024A&A...691A..52K,2025ApJ...980...36L}.
It is also proposed that the continuum shape is intrinsic, perhaps reflecting a binary massive black hole system or a two-component accretion-disk structure \citep[e.g.,][]{2025arXiv250505322I,2025arXiv251109278W,2025arXiv250512719Z}.
If we believe that V-shape continuum is a unique feature of a special population of AGN, narrow-line LRDs may represent a population of AGNs at the very low-mass end of the SMBH mass distribution, which could naturally lead to the absence of broad Balmer lines. The excesses in $\rm FWHM_{H\alpha}$ and $L_{\rm H\alpha}$ of the narrow-line LRDs compared to typical star-forming galaxies (Figure \ref{fig:Ha_OIII_poperties}) also suggest an additional contribution to the H$\alpha$ emission beyond that from star formation, possibly originating from a central AGN.

We therefore consider a scenario in which these narrow-line LRDs with V-shape continua are AGNs hosting low-mass SMBHs, leading to intrinsically narrow Balmer lines.
Within this framework, two possibilities may account for the non-detection of a classical broad component: (1) the broad line is sufficiently broad but intrinsically weak and thus buried below the current noise level; or (2) the broad-line width is intrinsically so narrow that it is blended into the observed line profile, analogous to narrow-line Seyfert~1 galaxies \citep[NLS1s; e.g.,][]{2004ApJ...606L..41G,2018MNRAS.480...96W}.
In the second scenario, the observed emission lines can arise from multiple components: the Balmer emission is partly contributed by an intrinsically narrow broad-line region (BLR) associated with a low-mass SMBH, while a conventional NLR is still present and produces narrower Balmer lines and forbidden lines such as \OIII. As a result, the observed H$\alpha$ emission can be enhanced relative to purely star-forming galaxies, which is consistent with the elevated H$\alpha$ luminosities and line widths observed in narrow-line LRDs compared to typical star-forming systems (Figure~\ref{fig:Ha_OIII_poperties}).
The second scenario is also in line with the recently proposed interpretation that the observed widths of the broad lines in high-redshift JWST AGN and LRDs are due to electron or Balmer scattering of intrinsic narrow lines by medium located outside the broad line regions rather than due to the motions of BLR clouds (\citealt{2025arXiv250316596N}; \citealt{2025arXiv250316595R}, but see \citealt{2025arXiv250403551J} for a debate). Narrow-line LRDs may represent a special population in which scattering is absent, allowing the intrinsic narrow line to be directly observed.

\subsubsection{Black Hole Mass Measurement}

Motivated by the two possibilities discussed above, we estimate conservative upper limits on the black hole masses of the narrow-line LRDs by considering both possibilities. First, since the majority of LRDs exhibit broad emission lines, it remains plausible that these sources also host a weak broad component hidden below the current noise level or within the observed narrow emission line. Indeed, joint analyses of PRISM and G395M data report broad-line evidence in three of our targets, where the detection are mainly driven by the PRISM data owing to its higher sensitivity \citep{Hviding2025}. However, the uncertainties in such joint analyses remain large because of the low spectral resolution of PRISM and the substantial difference in wavelength-dependent resolution between the grating and prism. If such broad components exist, they would imply very small BH masses. To evaluate the maximal strength of a potential broad component that could remain undetected at the present G395M spectra, we simulate the broad line detection probability of AGNs with different $M_{\rm BH}$ and $L_{\rm bol}$ under the current G395M noise level and observed narrow emission line.

We estimate the black hole masses of the narrow-line LRDs using the local virial relations, which link the black hole mass with the continuum/line luminosities and line widths. We use the relation calibrated by \citet{2013ApJ...775..116R}:
\begin{equation}
\begin{aligned}
&\log\left(\frac{M_{\rm BH,vir}}{\rm M_{\odot}}\right) = \log(\epsilon) + 6.57 \\
    &
    +0.47\log\left(\frac{{L_{\rm H\alpha}}}{10^{42} \rm ~erg~s^{-1}}\right)+2.06\log\left(\frac{{\rm FWHM}_{{\rm H} \alpha}}{10^3~\rm km~s^{-1}}\right),
\end{aligned}
\end{equation}
with a scaling factor $\epsilon=1.075$ following \citet{2015ApJ...813...82R}. We use this single-epoch BH mass calibration and the bolometric correction \citep{Chen2025} to link $L_{\rm H\alpha}$, $\rm FWHM_{\rm H\alpha}$, $M_{\rm BH}$, and $L_{\rm bol}$. Once $M_{\rm BH}$ and $L_{\rm bol}$ are specified, the corresponding broad \Ha\ component is fully determined. We use the best-fit single-Gaussian model of the observed \Ha\ line as the narrow component. We then combine the broad and narrow components, inject random noise sampled from the rms of the observed spectrum, and perform the two kinds of Gaussian fitting described in Section~\ref{subsec:emission_line_fitting}. For each $M_{\rm BH}$–$L_{\rm bol}$ pair, we estimate the broad-line detection probability using 50 simulations. A broad line is considered to be detected if the model with both narrow and broad components has a lower BIC than the model with only the narrow component.


\begin{figure*}
 \includegraphics[width=0.99\textwidth, trim=0cm 0cm 4cm 0cm, clip]{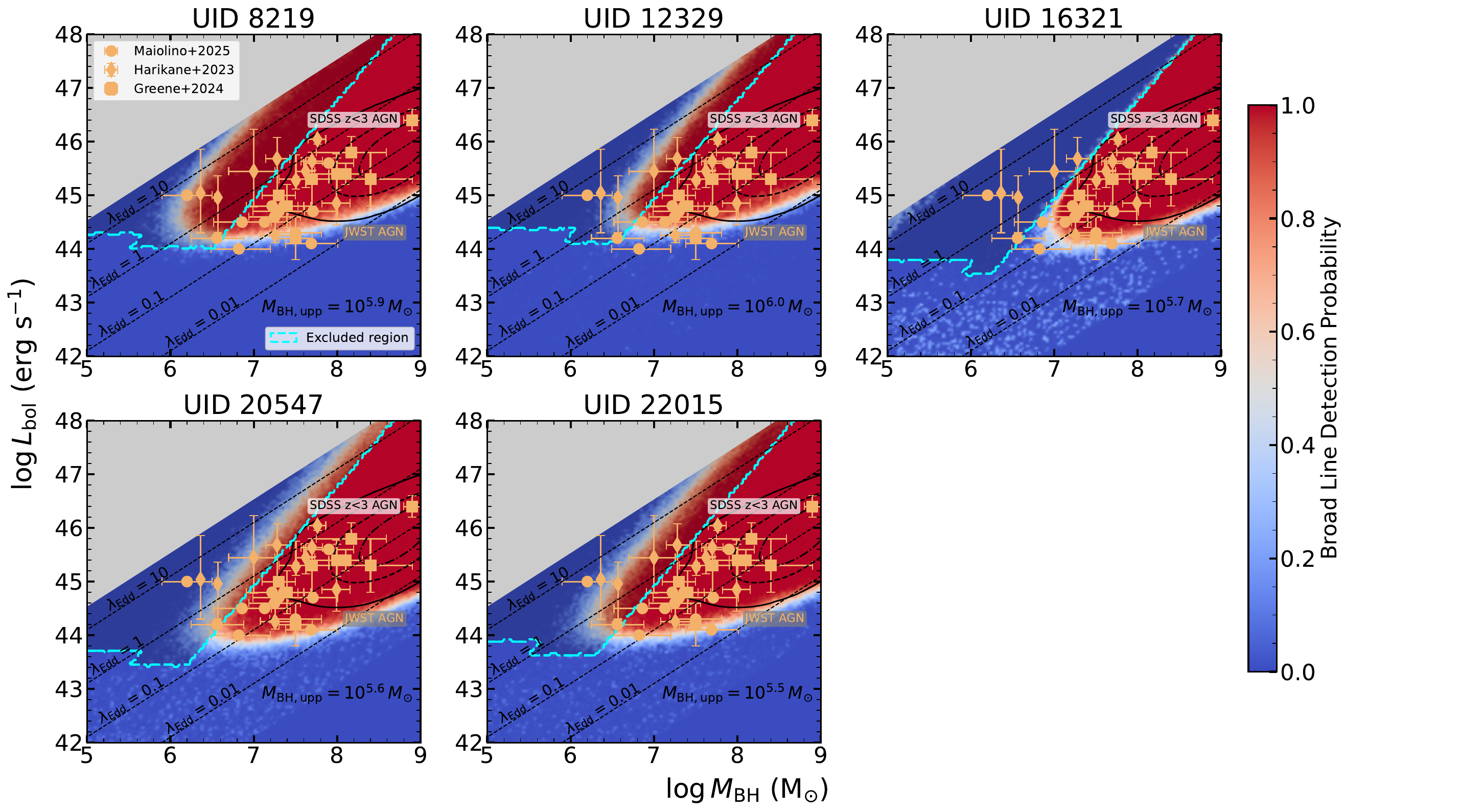}
 \centering
\caption{Broad-line detection probability of the five narrow-line LRDs in the $M_{\rm BH}$--$L_{\rm bol}$ space. The JWST AGNs \citep{2024ApJ...964...39G, 2024A&A...691A.145M, 2023ApJ...959...39H} and SDSS $z < 3$ AGNs \citep{Chen2025} are plotted for comparison. The dashed black lines indicate loci of constant Eddington ratio, $\lambda_{\rm Edd}=0.01, \,0.1,  \,1,  \,10$, respectively. The shaded region enclosed by cyan dashed line (very large $\lambda_{\rm Edd}$) indicates that the expected broad line can be significantly distinguished (broader/stronger) from the observed narrow emission line and therefore can be rule out. The black hole mass upper limit of each sources assuming $\lambda_{\rm Edd}=1$ is indicated in the bottom right of each panel. }
 \label{fig:BL_simulation}
\end{figure*}

The simulated broad-line detection probabilities of the five narrow-line LRDs are presented in Figure~\ref{fig:BL_simulation}. We also show the position of JWST AGNs and low-redshift quasars for comparison. The simulated result shows that the broad \Ha\ lines of these sources should already have been detected if their $M_{\rm BH}$ and $L_{\rm bol}$ are similar to other JWST AGNs. At high Eddington ratio ($\lambda_{\rm Edd}$) region, the FWHM of the expected broad line could be $<1000 ~\rm km~s^{-1}$, making them blended with the narrow line in some cases, similar to NLS1 in the local universe. In contrast, if $\lambda_{\rm Edd}$ is too low, the broad line could be very broad but too weak to be detected. The shaded region enclosed by the cyan dashed line (corresponding to large $\lambda_{\rm Edd}$) marks the parameter space where a broad \Ha\ component would be clearly distinguishable from the observed narrow emission line and can therefore be ruled out. In this regime, the simulated broad component becomes sufficiently broader or stronger than the narrow component. Operationally, we flag a model as distinguishable when either (1) the expected broad-line FWHM exceeds the narrow-line FWHM by more than $200~\mathrm{km\,s^{-1}}$ while remaining below $1000~\mathrm{km\,s^{-1}}$, and its flux is at least $40\%$ of the narrow component; or
(2) the broad-line FWHM is comparable to the narrow component (within $200~\mathrm{km\,s^{-1}}$), but its flux is larger than the narrow-line flux.
Therefore, the non-detection of broad line suggest that these narrow-line LRDs should have smaller $M_{\rm BH}$ and/or $\lambda_{\rm Edd}$ than other LRDs.

Assuming the Eddington ratio to be unity, the corresponding black hole mass of the ruled out region or the high broad-line detection probablity region can be regarded as an upper limit.
The resultant black hole mass upper limits are in the range of $\log_{10}(M_{\rm BH,uplim}/M_{\rm \odot}) = 5.5$–$6.0$.


For the second scenario, the entire observed H$\alpha$ line may arise from AGN activity if the BLR itself is intrinsically narrow. In this case, the observed line width can be treated as an upper limit on the intrinsic BLR velocity dispersion, while any contribution from the NLR or host galaxy would further reduce the true black hole mass.
Under this assumption, we apply the same virial relation using the H$\alpha$ luminosity and FWHM measured from the single-Gaussian fit. 
We therefore the use the intrinsic ${\rm FWHM}_{\rm H\alpha}$ and the total H$\alpha$ luminosity to estimate the black hole upper limit.
This procedure yields black hole mass upper limits of $\log_{10} (M_{\rm BH,uplim}/M_{\rm \odot}) =4.9$–$5.8$. These limits are conservative, as any host-galaxy contribution to H$\alpha$ would further lower the inferred black hole masses.

Because both scenarios are physically plausible and cannot be distinguished with the current data, we adopt the more conservative constraint by taking, for each source, the larger of the two upper limits derived from the two scenarios. The resulting black hole mass upper limits are all $\log_{10}(M_{\rm BH,uplim}/M_{\rm \odot}) \lesssim 6.0$, significantly smaller than those of typical JWST AGNs and broad-line LRDs.
We do not account for dust extinction in our simulations, so the inferred upper limits on black hole mass could be underestimated if significant extinction is present. 
Nevertheless, the impact of the luminosity correction on $M_{\rm BH}$ is modest compared to that of the line width. 
In addition, LRDs are expected to exhibit only modest dust extinction \citep{2025arXiv250518873C,2025arXiv250522600C,2025arXiv250302059S}. 
Even adopting a maximal value of $A_V = 1.5$ \citep[e.g.,][]{2025arXiv250522600C} would increase the black hole mass upper limits by only $\sim$0.3 dex.

\begin{figure}
 \includegraphics[width=0.5\textwidth]{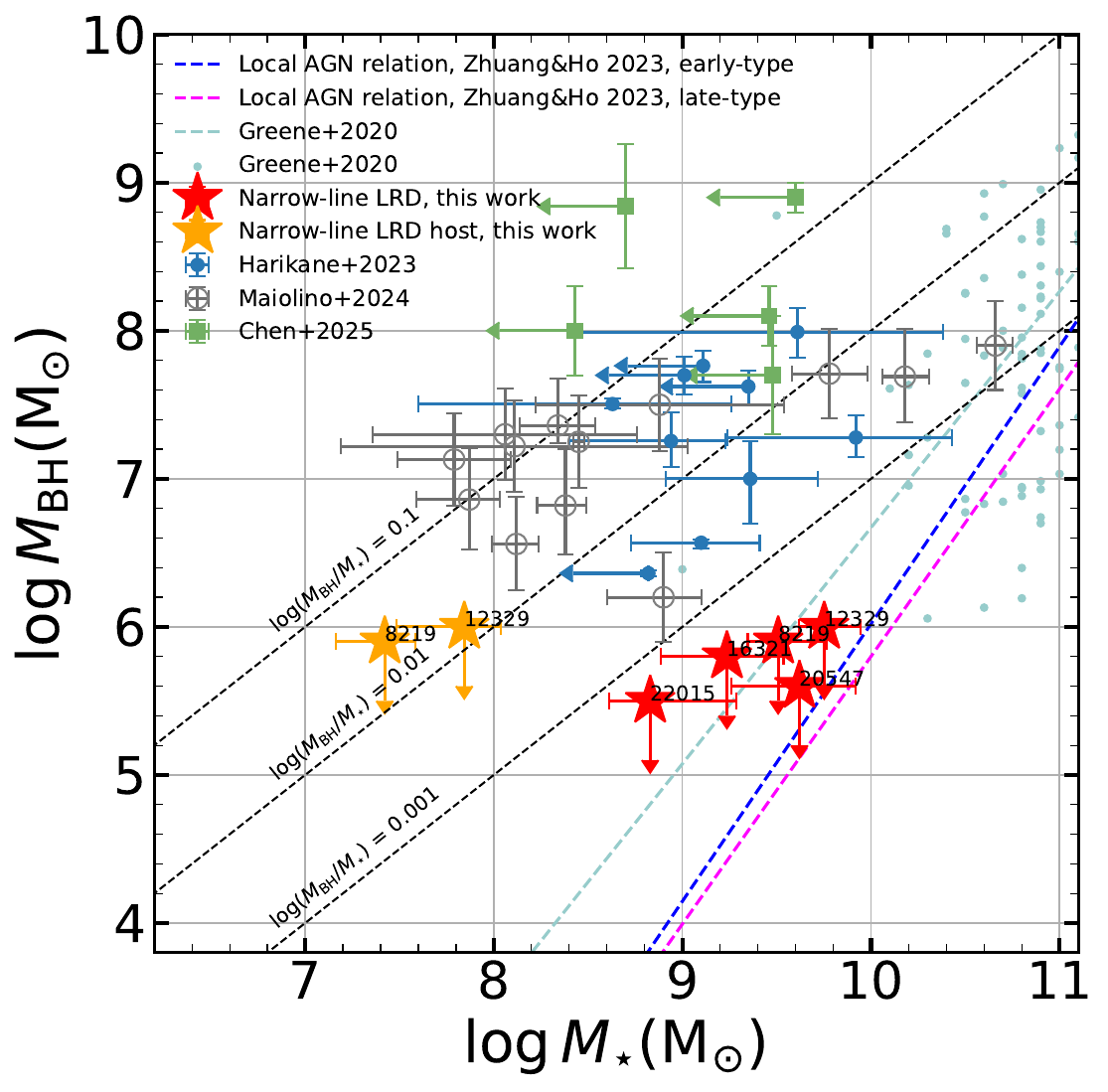}
 \centering
 \caption{Relation between black hole mass and the stellar mass of the host galaxy for the narrow-line LRDs. Measurements for the single-epoch black hole mass attribute the entire \Ha\ line from the AGN contribution. Red and orange stars represent cases where all the emission and only the extended emission, respectively, are attributed to the host galaxy. For comparison, we also include the high-redshift AGN samples from \citet[blue points]{2023ApJ...959...39H}, \citet[gray open circles]{2024A&A...691A.145M}, and \citet[light green squares]{2025ApJ...983...60C}. Colored dashed lines are the local relations for inactive \citep{Greene2020} and active \citep{2023NatAs...7.1376Z} galaxies. Black dashed lines represent different ratios of black hole mass to total stellar mass ($M_{\rm BH}/M_{*} =$ 0.001, 0.01 and 0.1).}
 \label{fig:MBH_Mstar}
\end{figure}

\subsubsection{$M_{\rm BH}$–$M_{\rm *}$ Relation}

We study these narrow-line LRDs in the context of the $M_{\rm BH}$–$M_{\rm *}$ relation. Due to the limited understanding of the unresolved central source, the results are open to multiple interpretations. The first possibility is that the emission from the unresolved point source mainly arises from extremely compact galaxies, as discovered by some works \citep[e.g.,][]{2023Natur.619..716C,2023ApJ...951...72O}. This is also consistent with the lack of significant AGN signatures in the narrow-line LRDs. We adopted the largest possible stellar masses estimated in Section \ref{subsec:pure_SF_gala}. The corresponding black hole mass versus stellar mass relation of the arrow-line LRDs is shown as the red stars in Figure \ref{fig:MBH_Mstar}. For comparison, we also plot the local ($z \approx 0$) scaling relation \citep{Greene2020,2023NatAs...7.1376Z} as well as other high-redshift Type 1 AGNs discovered with JWST \citep{2023ApJ...959...39H,2024A&A...691A.145M,2025ApJ...983...60C}. 
We find that four of the narrow-line LRDs are roughly consistent with or slightly below the local relation. The remaining narrow-line LRD likely hosts an overmassive black hole, but its $M_{\rm BH}$ is still smaller than those in other JWST Type 1 AGNs. 

Another possibility is that the central point sources are dominated by AGN, while the possible extended emission contributed by the host galaxies is weak. In this case, we adopt the morphological fitting result using a central point source and a \sersic\ component in Section \ref{subsec:emission_line_fitting}. Three narrow-line LRDs can be well fitted by a single PSF model. The other two (UID-8219 and UID-12329) exhibit obviously extended emission. For UID-12329, the extended emission is significantly off the center. Although such off-centered extended emission in LRDs may originate from purely nebular emission photoionized by the LRD itself \citep{2025arXiv250503183C}, we assume that it is from its host galaxy here. 

We perform SED modeling using \texttt{CIGALE} with the same configuration in Section \ref{subsubsec:star_mass_SFR_SEDmodeling} for the extended emission of UID-8219 and UID-12329. We add a relative error of 15\% in quadrature to the uncertainties of the fluxes to account for the uncertainties introduced by the image fitting. The best-fit SED result of UID-12329 is shown in Figure \ref{fig:cigale_12329}(b), with the SED of the PSF component shown for comparison. The derived stellar masses are substantially lower than those inferred assuming the entire emission arises from the host galaxy. The red rest-optical continuum is dominated by a central point source, whereas the extended emission appears significantly bluer. This color contrast is consistent with the results of \citet{2025ApJ...983...60C} for broad-line LRDs and suggests minimal dust extinction for the host galaxies, which in turn leads to lower stellar mass estimates. 
The results are shown as the yellow stars in Figure \ref{fig:MBH_Mstar}. 
In this case, the black hole masses are not as overmassive as those in other JWST AGNs but still lie above the local relations. If dust corrections are needed for $L_{\rm H\alpha}$, these objects would shift further away from the local scaling relations.

\subsection{X-ray Non-detection}
\begin{figure}
 \includegraphics[width=0.45\textwidth]{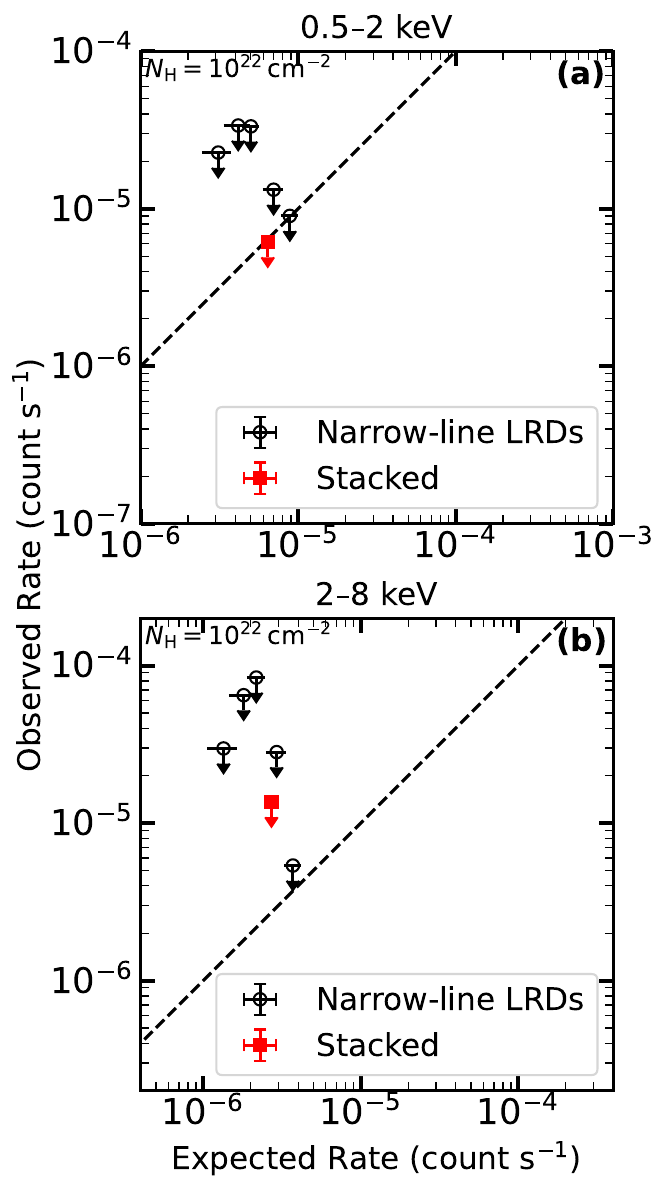}
 \centering
 \caption{Comparison of the observed upper limits of the count rate and the expected values from \Ha\ luminosities. Panel (a) and (b) show the 0.5--2 keV and 2--8 keV band, respectively. The black open circles and the red squares show the results of individual narrow-line LRDs and the stacking result, respectively. The dashed lines correspond to $x=y$ relation. The column densities are assumed to be $N_{\rm H} = 10^{22} \rm ~cm^{-2}$.}
 \label{fig:Xray_stacking}
\end{figure}

Previous X-ray stacking analyses of broad-line LRDs show that they have weak X-ray emission \citep[e.g.,][]{2024ApJ...969L..18A,2024ApJ...974L..26Y,2025MNRAS.538.1921M}, which may be attributed to radiative signatures of super-Eddington accreting black holes \citep[e.g.,][]{2024arXiv241000417M,2024arXiv241203653I}. We check the archival X-ray observations of the five narrow-line LRDs and find that they are all covered by the AEGIS-X Deep survey \citep{2015ApJS..220...10N} and the X-UDS survey \citep{2018ApJS..236...48K}, respectively. However, none of them has an individual X-ray detection. We use \texttt{CSTACK}\footnote{\url{https://lambic.astrosen.unam.mx/cstack/}.} and follow the method in \citet{2024ApJ...974L..26Y} to stack the X-ray observations and compare the results with the expected count rate using the $L_{\rm X} - L_{\rm H\alpha}$ relation in \citet{2012MNRAS.420.1825J}, assuming the column densities $N_{\rm H} = 10^{22} \rm ~cm^{-2}$. The stacking still yields a non-detection with a total exposure time of 2.2 Ms. The comparison of the observed upper limits of the count rates between the expected values from the observed \Ha\ luminosities for the 0.5--2 keV and 2--8 keV band is shown in Figure \ref{fig:Xray_stacking}. The soft band provides tighter constraints than the hard band. However, in both cases, the constraints for both the individual sources and the stacked sample are not stringent enough. The current depth of the X-ray data does not allow us to determine whether narrow-line LRDs are X-ray weak.

\section{Discussion}
\label{sec:discussion}
\subsection{Comparison with Other Works}
\label{subsec:sample_selection_compare}

We compare our LRD selection result with other studies, specifically \citet{2024ApJ...968...38K} and \citet{2024arXiv240403576K}. Cross-matching the compact sources in our parent sample with the \citet{2024ApJ...968...38K} LRD sample yields 38 common sources, of which 29 meet our LRD criteria. The remaining 9/38 ($\sim 24\%$) sources all exhibit very strong \Hb, \OIII, and \Ha\ emission lines, which boost their optical colors to be very red. After correcting for the emission line contribution (Section \ref{subsec:LRD_selection}), their rest-frame optical continuum slopes fall outside the LRD criteria. Similarly, cross-matching with \citet{2024arXiv240403576K} yields 61 common sources, with 50 classified as LRDs by our selection. Most of the remaining 11/61 ($\sim 18\%$) sources are also misclassified due to the strong emission lines, and two sources are due to the wrong photo-$z$.

\subsection{Velocity Dispersion and Host Dynamical Mass}
\label{subsec:sigma_Mdyn_relation}

The tight correlation between black hole mass and central (stellar) velocity dispersion $\sigma_{\rm stellar}$ has been extensively examined and firmly established in the local Universe \citep[e.g.,][]{Kormendy2013}. Spatially resolved observations leveraging gravitational lensing have extended investigations of this relationship to $z \sim 2$ \citep{2025arXiv250317478N}. At even higher redshifts, measurement of $\sigma_{\rm stellar}$ often uses spatially integrated gas emission lines as proxies due to the limit of spatial resolution \citep[e.g.,][]{2024A&A...691A.145M}. These high-redshift studies suggest that this relation is fundamental and may have little evolution. Assuming that narrow-line LRDs are AGNs hosting low-mass SMBHs, we investigate whether they follow these two relations.

\begin{figure*}
 \includegraphics[width=0.95\textwidth]{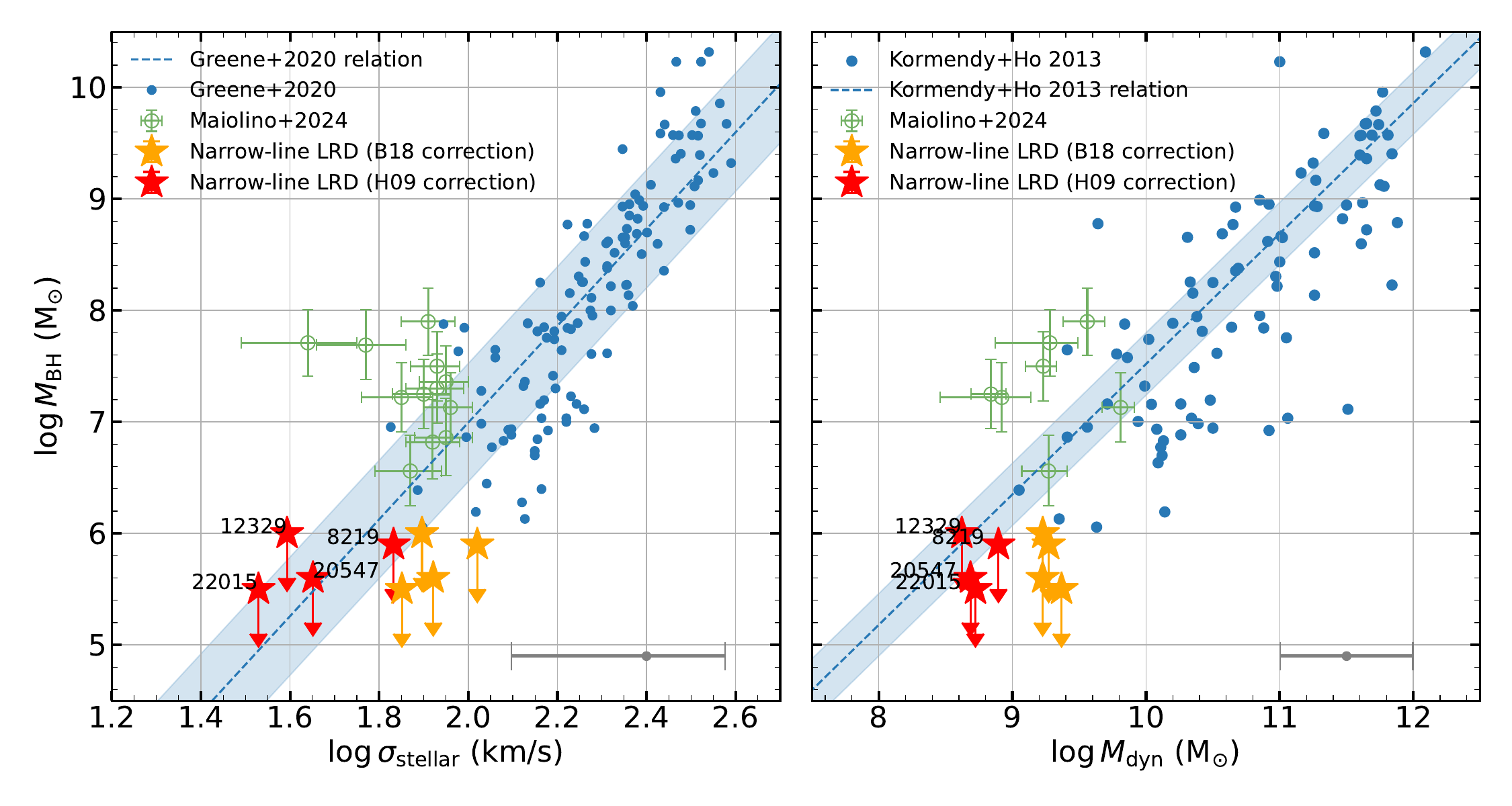}
 \centering
 \caption{Relations between black hole mass and the stellar velocity dispersion (left panel) and the dynamical mass of the host galaxy (right panel) for four narrow-line LRDs. Measurements for the single-epoch black hole mass attribute the entire \Ha\ line from the AGN contribution and are therefore represented as upper limits. The $\sigma_{\rm stellar}$ are converted from $\sigma_{\rm gas}$ measured from \OIII\ line using the relation in \citep[yellow stars]{2018ApJ...868L..36B} and \citep[red stars]{2009ApJ...699..638H}. For comparison, we also include the high-redshift AGN samples from \citep[green open dots]{2024A&A...691A.145M}. Blue dots and  dashed lines in the left and right panels are the local relations from \citet{Greene2020} and \citet{Kormendy2013}, respectively. The gray error bars in the lower right of each panel represent the median measurement errors of the x-axis.}
 \label{fig:MBH_Mdyn}
\end{figure*}

Four of the narrow-line LRDs (except UID-16321) have medium-resolution spectral coverage of the \OIII\ line, allowing us to measure the velocity dispersion of the ionizing gas ($\sigma_{\rm gas}$). As shown in Figure \ref{fig:Ha_OIII_poperties}(a),  $\rm FWHM_{H\alpha}$ is slightly larger than  $\rm FWHM_{[O\,\textsc{iii}]}$ for the narrow-line LRDs, possibly due to the contribution from the central AGN. Therefore, we measure $\sigma_{\rm gas}$ using the $\rm [O\,\textsc{iii}]$ line. The spectral resolution of the medium-grating for compact sources at the spectral range of interest is $80\sim 95 \rm~ km~s^{-1}$, which is comparable to the derived intrinsic velocity dispersion. Correcting for the LSF introduces a significant source of uncertainty, which dominates the final error budget. In reality, the integrated $\sigma_{\rm gas}$ measured using the emission line is not necessarily the same as the stellar velocity dispersion $\sigma_{\rm stellar}$ used in the local scaling relations. As done by \citet{2023A&A...677A.145U} and \citet{2024A&A...691A.145M}, we use the empirical relation between observed $\sigma_{\rm stellar}$ and $\sigma_{\rm gas}$ of low-redshift galaxies from \citet{2018ApJ...868L..36B} to correct the $\sigma_{\rm gas}$ and obtain a close estimate of $\sigma_{\rm stellar}$. However, the $\sigma_{\rm stellar}$--$\sigma_{\rm gas}$ relation in \citet{2018ApJ...868L..36B} is derived using massive galaxies, which may not be applicable to AGNs. Therefore, we also try to use the relation calibrated by \citet[Equation 3]{2009ApJ...699..638H} based on AGNs. These relations have a scatter of $\sim 0.2 $ dex that is not considered in our error budget.

Using the stellar velocity dispersion above and the morphological information from the PSF+\sersic\ fitting in Section \ref{subsec:morphology_image}, we estimate the dynamical masses of the narrow-line LRD host galaxies. Only UID-8219 and UID-12329 have significant extended emission detected, while the other two sources can be well fitted by a single PSF model. Therefore, the PSF+\sersic\ fitting result of UID-20547 and UID-22015 could have large uncertainty. Following \citet{2023A&A...677A.145U} and \citet{2024A&A...691A.145M} we adopt the equation:

\begin{equation}
    M_{\rm dyn} = K(n)K(q)\frac{\sigma_{\rm stellar}^2R_{\rm e}}{G},
\end{equation}
where $K(n) = 8.87 - 0.831n + 0.0241n^2$ and $n$ is the \sersic\ index \citep{2006MNRAS.366.1126C}, $K(q) = [0.87 +  0.38e^{-3.71(1-q)}]^2$ and $q$ is the axis ratio \citep{2022ApJ...936....9V}, and $R_{\rm e}$ is the effective radius. We use the $n$, $q$, and $R_{\rm e}$ of the \sersic\ component in the F150W or F200W band.

The estimated $\sigma_{\rm stellar}$ and $M_{\rm dyn}$ are reported in Table \ref{tab:stellar_BH_property}. The relations are shown in Figure \ref{fig:MBH_Mdyn} and compared with the high-redshift AGN samples from  \citet{2024A&A...691A.145M} and the local relation from \citet{Kormendy2013} and \citet{Greene2020}. For both the $M_{\rm BH}$–$\sigma_{\rm stellar}$ and $M_{\rm BH}$–$M_{\rm dyn}$ relations, the black holes appear undermassive relative to the local relation when adopting the \citet{2018ApJ...868L..36B} correction. Adopting the \citet{2009ApJ...699..638H} correction results in larger correction and therefore lower $\sigma_{\rm stellar}$ and $M_{\rm dyn}$ values. This may be because, in the AGN case, energy injected by AGN feedback accelerates the NLR gas to velocities exceeding the stellar virial velocities \citep{2009ApJ...699..638H}. Under this correction, the narrow-line LRDs could either be consistent with the local $M_{\rm BH}$–$\sigma_{\rm stellar}$ and $M_{\rm BH}$–$M_{\rm dyn}$ relations, or they may indeed host undermassive black holes. For UID-12329 and UID-20547, their stellar masses from the SED fitting assuming all emission is from galaxies are larger than the derived dynamical masses here (see Table \ref{tab:stellar_BH_property}), suggesting that the stellar masses are overestimated or the dynamical masses are underestimated. The black hole masses here are conservative upper limits. Therefore, in the AGN scenario, these narrow-line LRDs may represent a distinct population of AGNs that fall below the local $M_{\rm BH}$–$\sigma_{\rm stellar}$ and $M_{\rm BH}$–$M_{\rm dyn}$ relations, potentially indicating deviations from the idea that the $M_{\rm BH}$–$\sigma_{\rm stellar}$ relation is fundamental and already in place at high redshift. We notice that the current measurements suffer from large uncertainties. The expected $\sigma_{\rm stellar}$ for black holes with masses of $\sim 10^5~M_{\rm BH}$ is $< 40 \rm~km~s^{-1}$. Such a lower value can only be measured accurately by future higher resolution spectra.

\subsection{Implications of Narrow-line LRDs}
\label{subsec:implication}

The existence of narrow-line LRDs has critical implications for understanding the LRD population. Notably, narrow-line LRDs constitute 16\% of the V-shape continuum sources in our clean sample, a small but non-negligible fraction that challenges the AGN interpretation of this spectral feature.

If narrow-line LRDs are purely star-forming galaxies, their existence demonstrates that the V-shape SED can arise from compact starburst systems. This raises a key question: could broad-line LRDs also have significant host galaxy contributions to their continuum and emission lines? If so, the inferred AGN luminosities and black hole masses for broad-line LRDs may be systematically overestimated, as the host galaxy could dominate both the V-shape continuum and narrow-line emission.

Conversely, if narrow-line LRDs are AGNs, their low black hole masses ($10^5 \sim 10^6 ~ \rm M_{\odot}$) place them at the very low-mass end of the AGN population. These objects may represent an initial stage of the black hole growth, where accretion is super-Eddington but the black hole has yet to accumulate significant mass. The low fraction of narrow-line LRDs indicates that this evolutionary stage is short and the black hole grow very fast. As time goes on, their black holes may grow faster than the host galaxies, becoming more overmassive systems, resembling typical JWST AGNs \citep[e.g.,][]{2023ApJ...959...39H, 2024A&A...691A.145M, 2025ApJ...983...60C}.

Narrow-line LRDs may also represent the Type 2 counterparts of the broad-line LRDs, with the difference arising from torus obscuration. However, the striking similarity in SEDs between narrow-line and broad-line LRDs poses a puzzle. In local AGN populations, Type 1 and Type 2 sources typically exhibit distinct SEDs due to dust attenuation in the torus. If narrow-line LRDs are the Type 2 counterparts of broad-line LRDs, their SEDs should differ significantly. The observed similarity instead suggests either (1) a lack of a standard obscuring torus in LRDs \citep[e.g.,][]{2025ApJ...980...36L}, or (2) a scenario where both AGN and host galaxy contribute similarly to the V-shape continuum, decreasing orientation-dependent differences. The low fraction of narrow-line LRDs suggests that the BLR of LRDs have a large covering factor \citep[as suggested by][]{2025MNRAS.538.1921M} or their torus have a small opening angle. Future investigation of the near-IR and mid-IR properties of narrow-line LRDs will be essential to test these hypotheses. It is also worth noting that some Type 2 AGNs in the local universe intrinsically lack a BLR, as their X-ray spectra reveal no intrinsic absorption \citep[e.g.,][]{2012ApJ...759L..16H,2016A&A...594A..72P}. This scenario may potentially explain the similarity between the SED of narrow-line and broad-line LRDs, although it is difficult to confirm due to the lack of X-ray detections.

In either interpretation, narrow-line LRDs highlight the diversity within the LRD population and demand for a better method to decompose the AGN and galaxy contributions. Their existence challenges simplistic assumptions about the origin of V-shape SEDs and provides crucial insights into the formation pathways and duty cycles of LRDs.

\subsection{Relation between Broad \Ha\ Line and V-shape Continuum}
\label{subsec:relation_broadHa_Vshape}

Spectroscopic follow-up of photometrically identified LRDs reveal that $\gtrsim 70\%$ of them exhibit broad Balmer lines \citep[e.g.,][]{2023ApJ...952..142F,2024A&A...691A..52K,2024ApJ...964...39G,2024ApJ...963..129M,2024arXiv240302304W}. On the other hand, among AGNs with broad lines, only $\sim 30\%$ show a LRD-like V-shape SED \citep[e.g.,][]{2024arXiv240906772T,2025ApJ...979..138H,2025arXiv250403551J}. We further investigate the connection between the broad line emission and V-shape continuum. Specifically, we examine the distribution and fraction of sources with broad H$\alpha$ lines in the $\beta_{\rm UV}$–$\beta_{\rm opt}$ plane.

We use all sources with NIRSpec prism observation and high- or medium-resolution grating spectra covering \Ha. The prism spectra enable us to eliminate the emission line contribution when measure the continuum slope. We also require the sources to be compact, satisfying the same criterion described in Section~\ref{subsec:LRD_selection}. In total, there are 491 sources. These sources are classified into three categories: (1) AGNs with obvious broad \Ha\ line detections, defined by ${\rm BIC}_{\rm H\alpha2} + 6 < {\rm BIC}_{\rm H\alpha1}$; (2) sources without obvious broad \Ha\ line detections with ${\rm BIC}_{\rm H\alpha2} > {\rm BIC}_{\rm H\alpha1}$ + 1; and (3) the remaining sources labeled as `vague'. We visually inspect the fitting result to ensure that the broad \Ha\ line detections are real.

\begin{figure}
\hspace{-0.5cm}
 \includegraphics[width=0.5\textwidth]{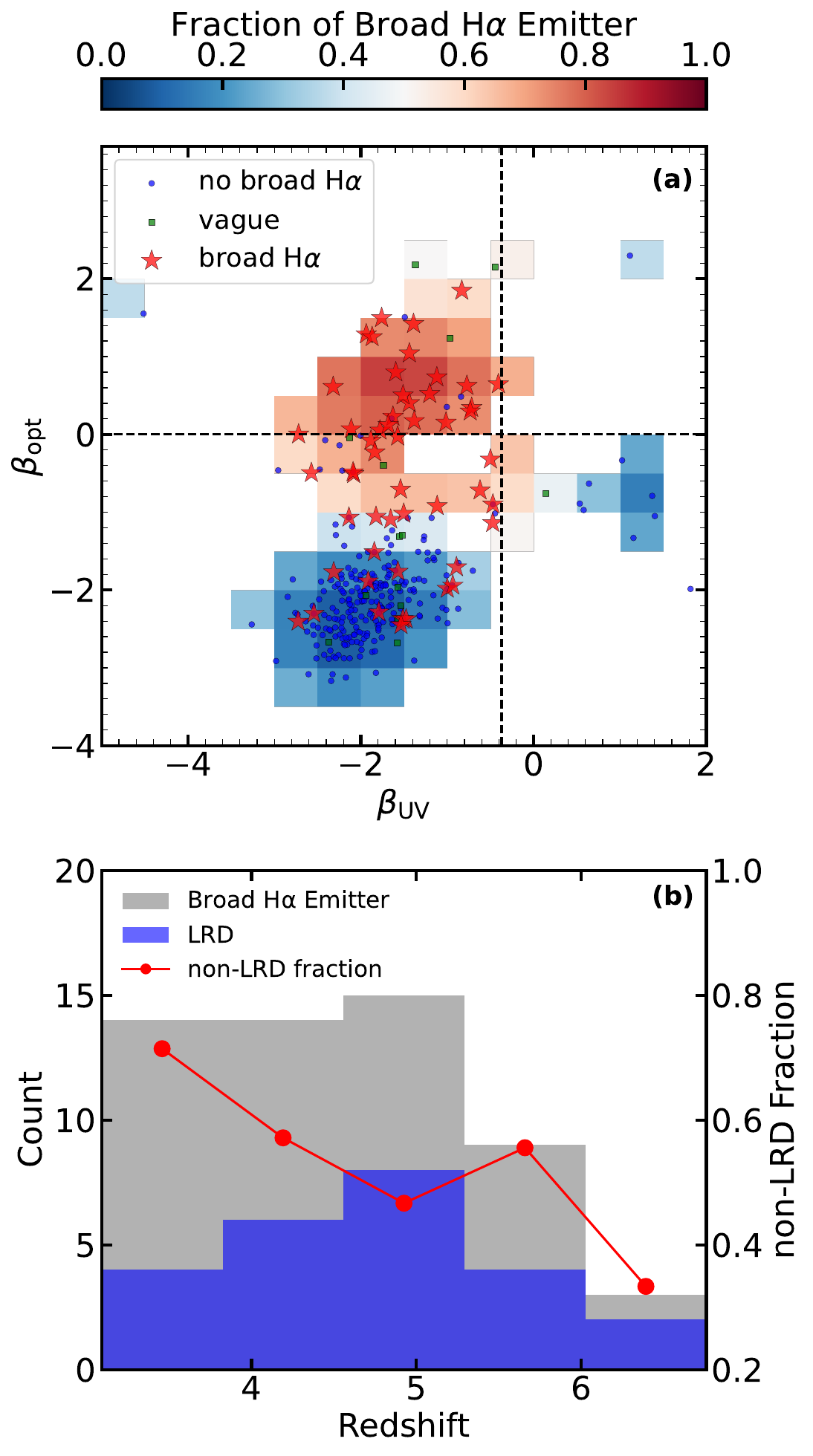}
 \centering
 \caption{(a) Rest-frame UV and optical slopes for sources with and without broad \Ha\ lines. The red stars and blue circles represent sources with and without broad \Ha\ lines, respectively. The green squares denote sources with uncertain detections. The color of the background squares represents the fraction of broad-line sources in each grid cell, calculated by assigning a weight of 1, 0.5, and 0 to the three types of sources, respectively. The fraction map is smoothed with a 2D Gaussian filter with a standard deviation of 1. (b) Redshift distribution of the broad \Ha\ lines emitters (gray) and broad-line LRDs (blue) in panel (a). Red curves shows the non-LRD fraction in each redshift bin.}
 \label{fig:broadHa_Vshape_relation}
\end{figure}

The $\beta_{\rm UV}$ versus $\beta_{\rm opt}$ distribution of these sources is shown in Figure \ref{fig:broadHa_Vshape_relation} (a). We divide the plane into 0.5$\times$0.5 cells and calculate the fraction of broad-line AGNs in each grid cell, where vague sources are counted as broad-line AGNs with a weight of 0.5. The resultant fraction map is smoothed using a 2D Gaussian filter with a standard deviation of 1. The distribution of the broad \Ha\ AGNs is similar to the result in \citet{2025ApJ...979..138H} , spanning a wide range of $\beta_{\rm opt}$. Compared to \citet{2025ApJ...979..138H} who compiled AGNs from various studies, our sample is selected in a uniform manner, and we also subtract the contribution of emission lines to accurately measure the intrinsic continuum slope. Among the 55 broad-line AGNs, 24 ($\sim 44\%$) of them exhibit V-shape continuum with $\beta_{\rm UV} < -0.37$ and $\beta_{\rm opt} > 0$. This fraction is slightly higher than that reported by previous studies \citep[e.g.,][]{2024arXiv240906772T,2025ApJ...979..138H,2025arXiv250403551J}, which may be due to the complex selection effects associated with the NIRSpec observations. The distribution of broad-line AGNs in the $\beta_{\rm UV}$–$\beta_{\rm opt}$ plane appears nearly uniform, suggesting that broad-line LRDs are likely AGNs located at the red end of the rest-frame optical slope distribution. In terms of the broad-line AGNs fraction, a clear demarcation appears at $\beta_{\rm opt} \sim -1$: on the redder side, the fraction of AGNs increases significantly and becomes dominant, while on the bluer side, the AGN fraction drops sharply due to the increasing number of star-forming galaxies. Therefore, selecting sources with red rest-frame optical colors can yield broad-line AGNs with high purity, which is in line with the result in \citet{2024ApJ...964...39G}. Such a selection is incomplete, as it misses a substantial number of AGNs located on the bluer end.

Another important question is the redshift evolution of the LRD fraction in AGNs, which would provide valuable insights into the growth and obscuration of black holes across cosmic time. In Figure \ref{fig:broadHa_Vshape_relation} (b) we show the redshift distributions of the broad \Ha\ lines emitters and broad-line LRDs, along with the corresponding non-LRD fractions in each redshift bin. The number of LRDs increases from high to intermediate redshift, then turns over and declines at $z \sim 5$, consistent with the results in \citet{2024arXiv240403576K}. The non-LRD fraction show a increasing trend toward low redshift, supporting the idea that LRDs represent an early phase of AGN evolution that is more prevalent at high redshift. This observed non-LRD fraction evolution is consistent with the predictions of \citet{2025arXiv250305537I}, who suggest that LRD correspond the very first phase of SMBH growth, and their fraction declines as SMBH undergone multiple accretion episodes over cosmic time.

\section{Conclusion}
\label{sec:conclusion}

In this paper, we have conducted a pilot investigation of narrow-line LRDs---sources characterized by unique V-shape continuum and compact morphology but lacking broad emission lines. Our study is based on all publicly available JWST NIRCAM and NIRSpec data in six JWST deep fields: A2744, CEERS, GOODS-S, GOODS-N, COSMOS, and UDS. We summarize our study as follows:

\begin{enumerate}
\item  We constructed a clean sample of 98 LRDs, selected based on V-shape continuum features derived from emission-line-free photometric slope fitting and compact morphology in the F444W band. Contamination from strong emission lines is effectively removed using spectroscopic information from NIRSpec/PRISM observations. Among the 32 LRDs with high- or medium-resolution NIRSpec grating spectra covering the \Ha\ region, we identified five ($\sim 16\%$) narrow-line LRDs that have no broad \Ha\ emission line.

\item We measured and compared the properties of narrow-line LRDs with those of broad-line LRDs and normal star-forming galaxies. LRDs tend to have larger $\rm FWHM_{H\alpha}$ and $L_{\rm H\alpha}$ compared with star-forming galaxies with similar $\rm FWHM_{[\rm O\,\textsc{iii}]}$ and $L_{[\rm O\,\textsc{iii}]}$. But narrow-line LRDs exhibit smaller \Ha\ EWs and larger \OIII\ EWs. Extended emission from two of the narrow-line LRDs is detected at short wavelength bands. 

\item If we assume all radiation of the narrow-line LRDs is from galaxy, our SED modeling reveals that they can be star-forming galaxies with high stellar masses ($\gtrsim 10^{9.5} \rm~M_{\odot}$) and SFRs ($\gtrsim 20 \rm~M_{\odot}~yr^{\rm -1}$). Narrow-line LRDs lie in the star-forming main sequence, but they have much higher dust attenuation compared with normal star-forming galaxies with similar \starmass\ and SFRs. The UV-based SFR and \Ha-based SFR are consistent in this scenario. This trend also holds for broad-line LRDs when using the narrow component of the \Ha\ line. Moreover, the gas $E(B-V)$ values derived from the Balmer Decrement for the narrow-line LRDs are consistent with those inferred from the SED modeling.

\item Narrow-line LRDs could also be a population of AGNs at the very low-mass end of the SMBH mass distribution. Assuming that the whole \Ha\ lines are from AGNs, the black hole masses are around $10^5 \sim 10^6 ~ \rm M_{\odot}$. These are very conservative upper limits with large uncertainties. Their black hole masses are smaller compared with other JWST Type 1 AGN, but they still likely host overmassive black holes compared with the local $M_{\rm BH}$–$M_{\rm *}$ relation. On the other hand, their black holes are consistent with (or undermassive) in terms of the $M_{\rm BH}$–$\sigma_{\rm stellar}$ and $M_{\rm BH}$–$M_{\rm dyn}$ relations. Narrow-line LRDs may represent an early stage of super-Eddington growth, where the black holes have yet to accumulate significant masses.

\item We also investigated the connection between broad lines and the V-shape continuum. We found that broad-line AGNs are broadly distributed in the $\beta_{\rm UV}$–$\beta_{\rm opt}$ plane, with the AGN fraction increasing sharply at $\beta_{\rm opt} \gtrsim -1$. While nearly half of the broad-line AGNs exhibit V-shape continua, many do not, suggesting that the V-shape is not a necessary feature of high-redshift broad-line AGNs. Our result shows that the red optical slope provides a high-purity but incomplete selection of AGNs. We also found an increasing LRD fraction among AGN towards higher redshift.
\end{enumerate}

In summary, our study reveals that narrow-line LRDs represent a distinct population. They may either be dusty, compact star-forming galaxies or AGNs with very low-mass black holes. Their spectral and structural properties provide new insights into the physical nature and diversity of LRDs. An unbiased study of flux-limited sample of narrow-line LRDs using deep slitless surveys \citep[e.g., COSMOS-3D, NEXUS, and SAPPHIRES;][]{2024jwst.prop.5893K,2024arXiv240812713S,2025arXiv250408039L,2025arXiv250315587S,2025arXiv250520393Z} will reduce selection effects and provide a better constraint on their fraction to improve our understanding of their role within the LRD population. Future deep JWST spectroscopy with higher resolution and sensitivity, along with deeper UV imaging, will be crucial for better understanding the nature of these sources. 

\begin{acknowledgments}

We acknowledge support from the National Key R\&D Program of China (2022YFF0503401), the National Science Foundation of China (12225301, 12233001), and the China Manned Space Program (CMS-CSST-2025-A09). 
Some of the data products presented herein were retrieved from the Dawn JWST Archive (DJA). DJA is an initiative of the Cosmic Dawn Center (DAWN), which is funded by the Danish National Research Foundation under grant DNRF140.
ZJZ thanks Changhao Chen and Shuqi Fu for helpful discussions.

\end{acknowledgments}

\facilities{JWST (NIRCam, NIRSpec)}

\software{\texttt{astropy} \citep{2013A&A...558A..33A,2018AJ....156..123A}, 
          \texttt{Bagpipe} \citep{2018MNRAS.480.4379C,2019MNRAS.490..417C},
          \texttt{CIGALE} \citep{2005MNRAS.360.1413B,2019A&A...622A.103B},
          \texttt{CSTACK},
          \texttt{GalfitM} \citep{2013MNRAS.430..330H,2013MNRAS.435..623V},
          \texttt{lmfit} \citep{2021zndo....598352N},
          \texttt{matplotlib} \citep{Hunter2007}, 
          \texttt{msafit} \citep{2024A&A...684A..87D}, 
          \texttt{numpy} \citep{Harris2020}, 
          \texttt{PSFEx} \citep{2011ASPC..442..435B},
          \texttt{PyPHER} \citep{2016ascl.soft09022B},
          \texttt{SExtractor} \citep{1996A&AS..117..393B},
          \texttt{TOPCAT} \citep{2005ASPC..347...29T},
          }

The data and code of this article are available on Zenodo under an open-source 
Creative Commons Attribution license: 
\dataset[doi:10.5281/zenodo.18149018]{https://doi.org/10.5281/zenodo.18149018}. The GitHub repository URL is \dataset[NL-LRD-2025-Data]{https://github.com/Zijian-astro/NL-LRD-2025-Data}.

\appendix

\section{Balmer Decrement}
\label{appendix:balmer_decrement}
Here we introduce the calculation of nebular extinction using Balmer Decrement. We derive the dust reddening following the extinction law of \citet{2000ApJ...533..682C}:

\begin{align}
E(B - V) &= \frac{E(\mathrm{H}\beta - \mathrm{H}\alpha)}{k(\lambda_{\mathrm{H}\beta}) - k(\lambda_{\mathrm{H}\alpha})} \\
&= \frac{2.5}{k(\lambda_{\mathrm{H}\beta}) - k(\lambda_{\mathrm{H}\alpha})} \log_{10} \left[ \frac{(\mathrm{H}\alpha/\mathrm{H}\beta)_{\mathrm{obs}}}{(\mathrm{H}\alpha/\mathrm{H}\beta)_{\mathrm{int}}} \right],
\end{align}
where $k(\lambda_{\rm H\beta}) =3.327$ and $k(\lambda_{\rm H\alpha}) = 4.598$ are the reddening curves evaluated at the \Hb\ and \Ha\ wavelengths, respectively. The factor $E(\mathrm{H}\beta - \mathrm{H}\alpha)$ is the color excess defined for \Hb\ and \Ha. $(\mathrm{H}\alpha/\mathrm{H}\beta)_{\mathrm{obs}}$ is the observed Balmer decrement and $(\mathrm{H}\alpha/\mathrm{H}\beta)_{\mathrm{int}}$ is the intrinsic Balmer decrement. We assume the value of $(\mathrm{H}\alpha/\mathrm{H}\beta)_{\mathrm{int}} = 2.86$, corresponding to a temperature $T = 10^4$ K and an electron density $n_e = 10^2 \rm~cm^{-3}$ for Case B recombination \citep{1989agna.book.....O}. We adopt the $E(B - V)$ values derived from the best-fit \texttt{CIGALE} model in Section \ref{subsec:pure_SF_gala}.


\bibliographystyle{aasjournalv7}
\bibliography{ms}{}

\end{document}